\documentclass[prd,aps,preprintnumbers,floats,floatfix,superscriptaddress,preprintnumbers,
showpacs,eqsecnum,nofootinbib,notitlepage,
twocolumn
]{revtex4-1}
\usepackage[utf8]{inputenc}
\usepackage{latexsym,array,theorem,mathrsfs,bm,float}
\usepackage{psfrag}
\usepackage{amsfonts,amsmath,amssymb,latexsym,array,afterpage,
theorem,mathrsfs,bm,float,epsfig,color,graphicx,tabularx,here,multirow}

\newcommand{\nn}{\nonumber \\}

\newcommand{\nablaA}{\overset{\scriptscriptstyle  \Gamma}{\nabla } {} }

\newcommand{\RA}{\overset{\scriptscriptstyle  \Gamma}{R } {} }

\newcommand{\Ta}{\overset{\scriptscriptstyle  (1)}{T } {} }
\newcommand{\Tb}{\overset{\scriptscriptstyle  (2)}{T } {} }
\newcommand{\Tc}{\overset{\scriptscriptstyle  (3)}{T } {} }

\newcommand{\Tcal}{ \mathcal{T} }
\newcommand{\Acal}{ \mathcal{A} }
\newcommand{\Bcal}{ \mathcal{B} }

\newcommand{\qV}{ \bm q }
\newcommand{\QV}{ \bm Q }
\newcommand{\pV}{ \bm p }
\newcommand{\Kcal}{ \mathcal{K} }
\newcommand{\Mcal}{ \mathcal{M} }
\newcommand{\Vcal}{ \mathcal{V} }
\begin{document}
\baselineskip=12pt

\preprint{YITP-19-67, IPMU19-0098}
\title{Ghostfree quadratic curvature theories with massive spin-2 and spin-0 particles}
\author{Katsuki \sc{Aoki}}
\email{katsuki.aoki@yukawa.kyoto-u.ac.jp}
\affiliation{Center for Gravitational Physics, Yukawa Institute for Theoretical Physics, Kyoto University, 606-8502, Kyoto, Japan}

\author{Shinji \sc{Mukohyama}}
\email{shinji.mukohyama@yukawa.kyoto-u.ac.jp}
\affiliation{Center for Gravitational Physics, Yukawa Institute for Theoretical Physics, Kyoto University, 606-8502, Kyoto, Japan}
\affiliation{Kavli Institute for the Physics and Mathematics of the Universe (WPI), The University of Tokyo, 277-8583, Chiba, Japan}

\date{\today}

 \begin{abstract}
  We consider generic derivative corrections to the Einstein gravity and find new classes of theories without ghost around the Minkowski background by means of an extension of the spacetime geometry. We assume the Riemann-Cartan geometry, i.e. a geometry with a non-vanishing torsion, and consider all possible terms in the Lagrangian up to scaling dimension four. We first clarify the number, spins, and parities of all particle species around the Minkowski background and find that some of those particle species are ghosts for generic choices of parameters. For special choices of the parameters, on the other hand, those would-be ghosts become infinitely heavy and thus can be removed from the physical content of particle species. Imposing the conditions on the coupling constants to eliminate the ghosts, we find new quadratic curvature theories which are ghost-free around the Minkowski background for a range of parameters. A key feature of these theories is that there exist a non-ghost massive spin-2 particle and a non-ghost massive spin-0 particle in the graviton propagator, as well as the massless spin-2 graviton. In the limit of the infinite mass of the torsion, the Riemann-Cartan geometry reduces to the Riemannian geometry and thus the physical content of particle species coincides with that of the well-known quadratic curvature theory in the metric formalism, i.e.~a massive spin-2 ghost, a massive spin-0 particle and the massless spin-2 graviton. Ghost-freedom therefore sets, besides other constraints, an upper bound on the mass of the torsion. In addition to the above mentioned particle species (a massive spin-2 particle, a massive spin-0 particle and the massless spin-2 graviton), the ghost-free theory contains either the set of a massive spin-1 and a massive spin-0 (Class I) or a couple of spin-1 (Class II). These additional particle species mediate gravity sourced by the spin of matter fields. 
 \end{abstract}



\maketitle

\section{Introduction}

Although Einstein's General Relativity (GR) is one of the most successful gravitational theories, it has been believed that GR is incomplete in the ultraviolet (UV) regime and is merely a low energy effective field theory (EFT). There are several attempts to unify quantum theory and gravity such as superstring theory, Ho\v{r}ava-Lifshiz gravity, ghost-free nonlocal gravity and so on. While constructing a complete quantum theory of gravity is certainly important, as a complementary attempt one should keep exploring phenomenological signatures of quantum gravity by means of a low energy EFT of gravity with quantum corrections to GR. One of such examples of this approach is Starobinsky model~\cite{Starobinsky:1980te}, where the Ricci scalar squared term is added to the Einstein-Hilbert action. This model yields a successful inflationary universe. Its predictions are in good agreement with the current observations~\cite{Akrami:2018odb} and future observations such as LiteBIRD~\cite{Hazumi:2019lys} can achieve the sensitivity sufficient to detect its inflationary prediction, the primordial gravitational waves.

A common feature of quantum gravity in the low energy regime is the appearance of derivative corrections to the Einstein-Hilbert action. The Ricci scalar squared mentioned above is indeed one of them. However, not only the Ricci scalar squared but also other generic higher curvature terms such as Ricci tensor squared are naturally expected to appear if there is no mechanism to prohibit them. Typical reasons to ignore them in the literature are for simplicity and due to the ghost problem. Except $f(R)$ theories, higher curvature terms in the action generally leads to higher derivatives of the graviton and then yields a massive spin-2 ghost as the Ostrogradsky ghost. If the ghost gives a dominant contribution, the system is outside the regime of validity of EFT and the theoretical control is lost. Therefore, in order for a theory to be predictive, the mass of the ghost (if exists) should be sufficiently heavier than the energy scale of interest, e.g. the inflation scale in the case of the inflationary universe. In particular, in the limit where the ghosts due to generic higher curvature terms are infinitely heavy, a higher curvature theory reduces to one of $f(R)$ theories, which can be recast to the form of scalar-tensor theories~\cite{Fujii:2003pa}. For this reason, it appears that phenomenologically interesting signatures of higher derivative corrections to the Einstein gravity come only from the spin-0 particle, at least in the context of the standard Riemannian geometry.

In the present paper, we point out that derivative corrections to the Einstein gravity can yield not only the spin-0 particle but also a massive spin-2 particle, a massive spin-1 particle and so on without suffering from light ghosts. The no-go argument in the previous paragraph is based on the Riemannian geometry, where the connection is computed by the metric and thus the metric is the only independent object that describes the spacetime geometry. From the first principle of geometry, on the other hand, the metric and the connection are in principle independent objects. We thus treat the metric and the connection as independent variables and study derivative corrections to the Einstein gravity. Imposing some conditions on the coupling constants of the theory, we find a new classes of theories in which the massless spin-2 graviton, the massive spin-2 particle and the massive spin-0 particle, which appear in the graviton propagator as in the usual quadratic curvature theory, can coexist without any ghost instability at least around the Minkowski background.

The construction of the new classes of higher curvature theories that we shall explore in the present paper is inspired by the recent developments of modified gravity, which had revealed that some of no-go results of ghost-free theories can be overturned. One of such examples is ghost-free massive gravity: although it was for a long time believed that a massive spin-2 field cannot be ghost-free beyond the linear order due to the Boulware-Deser ghost~\cite{Boulware:1973my}, the ghost-free nonlinear potential of the massive spin-2 field was found by de Rham et al.~in 2010~\cite{deRham:2010ik,deRham:2010kj}. Another example is ghost-free scalar-tensor theories with higher derivative terms: while it was for a long time supposed that equations of motion must be second order differential equations for a theory to avoid the Ostrogradsky ghost, the discovery of the degenerate higher order scalar-tensor (DHOST) theories~\cite{Gleyzes:2014dya,Gleyzes:2014qga,Zumalacarregui:2013pma,Langlois:2015cwa,Langlois:2015skt,Crisostomi:2016czh,Achour:2016rkg,BenAchour:2016fzp} opened up a new window on constructing ghost-free higher order theories via degeneracy of the kinetic matrix. Therefore, we consider it worthwhile to revisit higher curvature theories in light of these recent developments, in particular by combining the ideas behind these recent successes and seeking possible new signatures of higher curvature corrections to the Einstein gravity.

The rest of the present paper is organized as follows. In Section \ref{sec_action}, we first give a general discussion on a gravitational Lagrangian receiving derivative corrections in the framework beyond the Riemannian geometry. We then assume the metric compatibility condition just for simplicity and provide the explicit Lagrangian relevant to the perturbation analysis around the Minkowski background. In Section \ref{sec_Lorentz_inv}, we compute the quadratic Lagrangian around the Minkowski background and show the kinetic structure of the theory. The number, spins, and parities of particle species in the theory are then identified. Introducing a matter field, we also compute the graviton propagator of the theory. The main result is shown in Section \ref{sec_stability}, where the stability (no ghost, no tachyon) conditions around the Minkowski background are derived based on the $3+1$ decomposition. The stability conditions and the particle contents of the ghost-free theories are summarized in \S.~\ref{subsec_summary}. Finally, we give concluding remarks in Section \ref{summary}.

\section{Action}
\label{sec_action}
\subsection{General discussion}
It is known that quantum corrections give rise to derivative corrections to the Einstein-Hilbert action,
\begin{align}
\mathcal{L}=\mathcal{L}(g,\partial g,\partial^2 g)=\mathcal{L}_R+\mathcal{L}_{R^2}+\cdots\,, \label{quadratic_gravity}
\end{align}
where
\begin{align}
\mathcal{L}_R&=\frac{M_{\rm pl}^2}{2} R\,, \\
\mathcal{L}_{R^2}&=\alpha_1 R^2 +\alpha_2 R_{\mu\nu}R^{\mu\nu}+\alpha_3 R_{\mu\nu\rho\sigma}R^{\mu\nu\rho\sigma}\,,
\end{align}
and $\cdots$ in \eqref{quadratic_gravity} represent terms involving more than four derivatives acted on the metric. Here, we have set the cosmological constant to zero in order to admit the Minkowski background solution. In four dimension, one of the parameters $\alpha_i$ can be set to zero by means of the Gauss-Bonnet theorem. From now on, we thus set $\alpha_3=0$ so that the theory is parameterized by the two parameters $(\alpha_1,\alpha_2)$ in addition to the Planck mass. If terms in $\cdots$ in \eqref{quadratic_gravity} are suppressed by $M_{\rm pl}^2$ then one may use \eqref{quadratic_gravity} to discuss a classical dynamics of gravity as long as $|\nabla_{\alpha_1}\cdots\nabla_{\alpha_n} R_{\mu\nu\rho\sigma}| \ll M_{\rm pl}^{n+2}$ for all $n\geq 0$ in tetrad basis. From naturalness it is then expected that $|\alpha_{1,2}|=\mathcal{O}(1)$. Nonetheless, in some cases $|\alpha_1|$ or/and $|\alpha_2|$ might be parametrically larger than $\mathcal{O}(1)$ for some reasons. Indeed, actual values of $|\alpha_1|$ and $|\alpha_2|$ depend on the UV completion of GR, i.e. quantum gravity. In the lack of complete understanding of quantum gravity, we should not a priori exclude the possibility of large $|\alpha_1|$ or/and $|\alpha_2|$. If $|\alpha_1| \gg 1$ or/and $|\alpha_2| \gg 1$ and if terms in $\cdots$ of \eqref{quadratic_gravity} are still suppressed by $M_{\rm pl}^2$ then the quadratic curvature corrections can give relevant contributions even in the classical regime below $M_{\rm pl}$ and can give various interesting phenomenology such as Starobinsky inflation~\cite{Starobinsky:1980te}. However, it is known that the theory \eqref{quadratic_gravity} generically has a massive spin-2 ghost in addition to a massive spin-0 degree and the massless spin-2 graviton.

A possible solution to the spin-2 ghost problem is to assume the hierarchy $|\alpha_1| \gg |\alpha_2|$ (with $\alpha_3=0$) as usually postulated in the literature so that the mass of the spin-2 ghost is much heavier than that of the spin-0 degree. In this case, one can safely integrate out the massive spin-2 ghost in the regime of validity of the EFT describing the physics of the spin-0 degree. In practice, if we assume $\alpha_2=0$ then the massive spin-2 ghost is infinitely heavy and the gravitational sector of the low energy EFT has the massless spin-2 graviton and the massive spin-0 degree only. In the present paper, on the other hand, we shall deform the theory so that the massive spin-2 mode becomes non-ghost while keeping the stable massive spin-0 mode and the massless spin-2 graviton. A new ingredient in the present paper that makes such deformation possible is an extension of the spacetime geometry.

We shall treat the metric and the connection as independent variables. The geometry with a metric and a generic affine connection is called the metric-affine geometry. In the low energy regime, it would be sufficient to include terms up to scaling dimension four. The Lagrangian is generally given by
\begin{align}
\mathcal{L}_G=\mathcal{L}_G(g,\partial g,\partial^2 g, \Gamma,  \partial \Gamma),
\end{align}
where the scaling dimensions of the metric and the connection are supposed to be $[g_{\alpha\beta}]=0$ and $[\Gamma^{\alpha}_{\beta\gamma}]=1$, respectively. Since the partial derivative is not a covariant quantity, the diffeomorphism invariance then leads to the requirement that the Lagrangian should be built out of geometrical quantities and their covariant derivatives as
\begin{align}
\mathcal{L}_G=\mathcal{L}_G(g,\RA,T,Q,\nablaA T ,\nablaA Q), 
\end{align}
where
\begin{align}
\RA{}^{ \mu}{}_{\nu\alpha\beta}(\Gamma)&:=\partial_{\alpha}\Gamma^{\mu}_{\nu\beta}-\partial_{\beta}\Gamma^{\mu}_{\nu\alpha}
+\Gamma^{\mu}_{\sigma\alpha } \Gamma^{\sigma}_{\nu\beta }-\Gamma^{\mu}_{ \sigma \beta} \Gamma^{\sigma}_{\nu\alpha } 
,
\\
T^{\mu}{}_{\alpha\beta}&:=\Gamma^{\mu}_{\beta\alpha}-\Gamma^{\mu}_{\alpha\beta}
,\\
Q_{\mu}{}^{\alpha\beta}&:=\nablaA{}_{\mu}g^{\alpha\beta}, 
\end{align}
are the curvature, the torsion, and the non-metricity tensors, respectively. The covariant derivative of a vector is defined as
\begin{align}
\nablaA_{\mu}A^{\nu}=\partial_{\mu}A^{\nu}+\Gamma^{\nu}_{\rho\mu}A^{\rho} .
\end{align}

We suppose that all gravitational degrees of freedom except the standard massless spin-2 graviton are massive so that the fifth force, existence of which is strongly constrained by the solar system experiments~(see \cite{Will:2014kxa} for example), does not appear in the low energy limit. In this case, it is useful to decompose the general affine connection $\Gamma^{\mu}_{\alpha\beta}$ into the Riemannian part and the deviations from it, namely the Levi-Civita connection $\left\{ {}^{\,\, \mu}_{\alpha\beta} \right\}$ and the distortion tensor $\kappa^{\mu}{}_{\alpha\beta}$:
\begin{align}
\Gamma^{\mu}{}_{\alpha\beta}=
\left\{ {}^{\,\, \mu}_{\alpha\beta} \right\}
+\kappa^{\mu}{}_{\alpha\beta}
.
\end{align}
The torsion and the non-metricity are then expressed by
\begin{align}
T^{\mu}{}_{\alpha\beta}=2\kappa^{\mu}{}_{[\beta\alpha]}
, \quad
Q_{\mu}{}^{\alpha\beta}=2\kappa^{(\alpha\beta)}{}_{\mu}
,
\end{align}
or, conversely, the distortion is given by
\begin{align}
\kappa^{\mu}{}_{\alpha\beta}&=
-\frac{1}{2}(T^{\mu}{}_{\alpha\beta}-T_{\beta}{}^{\mu}{}_{\alpha}+T_{\alpha\beta}{}^{\mu} )
\nn
&
-\frac{1}{2}(Q^{\mu}{}_{\alpha\beta}-Q_{\beta}{}^{\mu}{}_{\alpha}-Q_{\alpha\beta}{}^{\mu} )
,
\end{align}
Hence, the Lagrangian can be rewritten as
\begin{align}
\mathcal{L}_G=\mathcal{L}_G(g,R,T,Q,\nabla T ,\nabla Q)
, \label{action}
\end{align}
where $R^{\mu}{}_{\alpha\beta\gamma}$ and $\nabla_{\mu}$ are the Riemann curvature and the covariant derivatives defined by the Levi-Civita connection. In the form \eqref{action}, we can regard the torsion tensor and the non-metricity tensor as independent variables instead of parts of the connection. In energy scales below the masses of the torsion and the non-metricity, all components of them may be integrated out; the gravitational theory is then represented by the metric only. Therefore, in the context of particle physics, what we did here is introducing new heavy particles in UV regime (see \cite{Percacci:2009ij} for more details on this point).


\subsection{Metric compatible theory}

Hereafter, we impose the metric compatibility condition
\begin{align}
Q_{\mu}{}^{\alpha\beta}=0,
\end{align}
just for simplicity. The geometry with a metric and a metric compatible connection is then called Riemann-Cartan geometry. This geometry particularly has gained attention in the literature since one can choose the tetrad $e^a{}_{\mu}$ and the anti-symmetric spin connection $\omega^{ab}{}_{\mu}=\omega^{[ab]}{}_{\mu}$, where $a,b,\cdots$ are the Lorentz indices, as independent variables and can regard them as gauge fields associated with the local translation and the local rotation, respectively. In this picture, the curvature two-form $\frac{1}{2}\RA{}^{ab}{}_{\mu\nu}dx^{\mu}dx^{\nu}$ and the torsion two-form $\frac{1}{2}T^a{}_{\mu\nu}dx^{\mu}dx^{\nu}$ are interpreted as the field strengths of them (see~\cite{Blagojevic:2002du,Obukhov:2018bmf} for reviews). Theories whose Lagrangian is algebraically constructed by the curvature and the torsion are called the Poincar\'{e} gauge theories (PGTs). On the other hand, in the present paper we shall include not only all the ingredients used in the Lagrangian of PGTs but also terms quadratic in derivatives of the torsion in the Lagrangian because there would be no reason to exclude them since both the curvature squared and the torsion derivative squared have scaling dimension four.

We assume that the cosmological constant vanishes to admit the Minkowski spacetime as a vacuum solution and study linear perturbations around the Minkowski background. The general parity preserving Lagrangian up to scaling dimension four is then given by the following pieces
\begin{align}
\mathcal{L}_G&=\mathcal{L}_R +\mathcal{L}_{T^2} +\mathcal{L}_{R^2}+ \mathcal{L}_{(\nabla T)^2}+\mathcal{L}_{R \nabla T} 
\nn
&+
\mathcal{L}_{R T^2}+
\mathcal{L}_{(\nabla T) T^2} + \mathcal{L}_{T^4} . \label{LG}
\end{align}
The last three parts, which include terms schematically represented by the suffixes, are irrelevant to linear perturbation analysis around the Minkowski spacetime. By the use of the equivalence upon integration by parts and the Bianchi identity, the most general expressions of the relevant terms are 
\begin{align}
\mathcal{L}_R&=\frac{M_{\rm pl}^2}{2} R
, \\
\mathcal{L}_{R^2}&=\alpha_1 R^2 +\alpha_2 R_{\mu\nu}R^{\mu\nu}+\alpha_3 R_{\mu\nu\rho\sigma}R^{\mu\nu\rho\sigma}
, \\
\mathcal{L}_{R\nabla T}
&=\beta_1 R_{\mu\nu} \nabla_{\rho}\Ta^{\mu\nu\rho}+\beta_2 R \nabla_{\mu} T^{\mu}
, \\
\mathcal{L}_{T^2}&=\frac{M_T^2}{2}(a_1 \Ta^{\mu\nu\rho}\Ta_{\mu\nu\rho}+a_2  T^{\mu}T_{\mu}+a_3\mathcal{T}^{\mu} \mathcal{T}_{\nu})
, \\
\mathcal{L}_{(\nabla T)^2}&=b_1 \nabla_{\mu} \Ta_{\nu\rho\sigma}\nabla^{\mu} \Ta^{\nu\rho\sigma}+b_2 \nabla_{\mu} \Ta^{\mu\rho\sigma} \nabla_{\nu} \Ta^{\nu}{}_{\rho\sigma} 
\nn
&+ b_3 \nabla_{\mu} \Ta^{\rho\sigma\mu} \nabla_{\nu} \Ta_{\rho\sigma}{}^{\nu}
+b_4 \nabla_{\mu}T_{\nu} \nabla^{\mu} T^{\nu}
\nn
&
+b_5 \nabla_{\mu} T^{\mu} \nabla_{\nu} T^{\nu}
+b_6 \nabla_{\mu}\Tcal_{\nu} \nabla^{\mu} \Tcal^{\nu}  +b_7 \nabla_{\mu} \Tcal^{\mu} \nabla_{\nu} \Tcal^{\nu}
\nn
&+b_8\nabla_{\mu}\Ta^{\mu\nu\rho}\nabla_{\nu}T_{\rho} 
+b_9 \epsilon^{\mu\nu\rho\sigma} \nabla_{\alpha} \Ta^{\alpha}{}_{\mu\nu} \nabla_{\rho}\Tcal_{\sigma}
\label{L_nablaT}
,
\end{align}
where $\alpha_i,\beta_i,a_i,$ and $b_i$ are dimensionless parameters. The torsion tensor has been decomposed into three irreducible pieces,
\begin{align}
\Tb_{\mu\nu\rho} &=\frac{2}{3}g_{\mu[\nu}T_{\rho]}
, \\
\Tc_{\mu\nu\rho}&=\epsilon_{\mu\nu\rho\sigma}\mathcal{T}^{\sigma} \label{T3}
, \\
\Ta_{\mu\nu\rho}&=T_{\mu\nu\rho}-\Tb_{\mu\nu\rho}-\Tc_{\mu\nu\rho}
\end{align}
with
\begin{align}
T_{\mu}&:=T^{\nu}{}_{\nu\mu}
, \\
\mathcal{T}_{\mu}&:=\frac{1}{6}\epsilon_{\mu\nu\rho\sigma}T^{\nu\rho\sigma}
.
\end{align}
The irreducible piece $\Ta_{\mu\nu\rho}$ satisfies
\begin{align}
\Ta_{\mu(\nu\rho)}=0,\quad \Ta_{[\mu\nu\rho]}=0, \quad \Ta^{\mu}{}_{\mu\nu}=0
\end{align}
and thus has $16$ independent components. As we have already stated, in light of the Gauss-Bonnet theorem we set $\alpha_3=0$ without loss of generality. Then, the remaining 16 dimensionless parameters $(\alpha_i,\beta_i,a_i,b_i)$ characterize linear perturbations in addition to the two mass parameters $M_{\rm pl}$ and $M_T$. The terms $\mathcal{L}_{T^2}$ act as the mass terms of the torsion. Hence, we assume $a_i\neq 0$ throughout in order to recover GR in the low energy limit.

There are a few general remarks on the Lagrangian \eqref{LG}. (i) Solutions of \eqref{LG} generally have a non-vanishing torsion because $\mathcal{L}_{R\nabla T}$ gives source terms of the torsion equation of motion. Since the source terms are proportional to
\begin{align}
\nabla_{\rho}R_{\mu\nu},~ \nabla_{\mu} R
,
\end{align}
an Einstein manifold $R_{\mu\nu}=\frac{R_0}{4}g_{\mu\nu}$ with a constant $R_0$ can be a torsionless solution. In particular, Ricci flat spacetimes with vanishing torsion are vacuum solutions of \eqref{LG}. (ii) In the case of $\beta_1=\beta_2=0$, the metric perturbations and the torsion perturbations are decoupled from each other at linear order around the Minkowski background. This case is less interesting because the massive spin-2 ghost exists as long as $\alpha_2 \neq 0$. (iii) When we take the limit $M_T \rightarrow \infty$ with finite $\beta_i$ and $b_i$, all torsional degrees of freedom become infinitely heavy and thus can be integrated out, and the theory is represented by the metric only. Then, the Lagrangian \eqref{quadratic_gravity} is obtained as an effective theory.

Before finishing this section, it would be worth mentioning the relation to the parity preserving quadratic PGT
\begin{widetext}
\begin{align}
\mathcal{L}_{\rm PGT^+}&=\lambda \RA + (r_4+r_5) \RA_{\mu\nu} \RA^{\mu\nu} + (r_4-r_5) \RA_{\mu\nu} \RA^{\nu\mu} +
\left( \frac{r_1}{3}+\frac{r_2}{6} \right) \RA^{\mu\nu\rho\sigma} \RA_{\mu\nu\rho\sigma} 
\nn
& 
+\left( \frac{2r_1}{3}-\frac{2r_2}{3}\right) \RA^{\mu\nu\rho\sigma} \RA_{\mu\rho\nu\sigma} 
+\left( \frac{r_1}{3}+\frac{r_2}{6}-r_3 \right) \RA^{\mu\nu\rho\sigma} \RA_{\rho\sigma\mu\nu} 
\nn
&+\left( \frac{\lambda}{4}+\frac{t_1}{3}+\frac{t_2}{12}\right) T^{\mu\nu\rho}T_{\mu\nu\rho} 
+\left( -\frac{\lambda}{2}-\frac{t_1}{3}+\frac{t_2}{6} \right) T^{\mu\nu\rho}T_{\nu\rho\mu} 
+ \left( -\lambda-\frac{t_1}{3}+\frac{2t_3}{3}\right)T^{\mu}{}_{\mu\rho}T^{\nu}{}_{\nu\rho} , \label{PGT}
\end{align}
where the term $\RA^2$ is omitted by the use of the Gauss-Bonnet theorem. We follow the parameterization used in \cite{Sezgin:1979zf,Lin:2018awc}. The Lagrangian has 8 parameters in addition to the gravitational constant $\lambda$. This Lagrangian contains up to the first derivative of the tetrad and the anti-symmetric spin connection, while the previous Lagrangian \eqref{LG} contains the second derivative of the tetrad due to the presence of the derivatives of the torsion.
Therefore, PGTs are clearly a subset of the general Lagrangian \eqref{LG}. Indeed, \eqref{PGT} is obtained from \eqref{LG} by setting
\begin{align}
\lambda&=\frac{M_{\rm pl}^2}{2}
, ~ 
\alpha_1=-r_1+r_3 , ~ \alpha_2 =2 (2r_1-2r_3+r_4) 
, ~
\beta_1=-4(2r_1-2r_3+r_4) , ~ \beta_2=-\frac{4}{3}(r_1-r_3+2r_4), 
\nn
a_1&=\frac{1}{M_T^2}(t_1+2\lambda) , ~ a_2=\frac{4}{3M_T^2}(t_3-2\lambda) , ~ a_3=-\frac{3}{M_T^2}(t_2-2\lambda), 
\nn 
b_1&=r_1 , ~b_2=\frac{1}{2}(-4r_1+4r_3-r_4+r_5) , ~b_3=2(r_1-2r_3+r_4), 
\nn
b_4&=\frac{4}{9}(r_1+r_4+r_5) ,~ b_5=\frac{4}{9}(2r_1-3r_3+5r_4-r_5)
, ~ b_6=-2r_3-r_5 , ~ b_7=-\frac{3}{2}r_2+2r_3+r_5, 
\nn
b_8&=-\frac{4}{3}(r_1+r_4+r_5), ~ b_{9}=2r_3+r_5, \label{PGTpara} 
\end{align}
\end{widetext}
and appropriately choosing the remaining Lagrangian $\mathcal{L}_{RT^2},\mathcal{L}_{(\nabla T) T^2},$ and $\mathcal{L}_{T^4}$, where we have used the Gauss-Bonnet theorem to eliminate the Riemann squared term.


\section{Gravitational degrees of freedom and critical conditions}
\label{sec_Lorentz_inv}
\subsection{Quadratic Lagrangian}
We study perturbations around the Minkowski background
\begin{align}
g_{\mu\nu}=\eta_{\mu\nu}+\delta g_{\mu\nu},
\end{align}
where we consider the torsion itself as a perturbation quantity since the background is torsionless. The metric perturbation can be decomposed into 
\begin{align}
\delta g_{\mu\nu}=h^{TT}_{\mu\nu}+2\partial_{(\mu}h^T_{\nu)}+\left(\partial_{\mu}\partial_{\nu}-\frac{1}{4}\eta_{\mu\nu}\Box \right) \sigma +\frac{1}{4}\eta_{\mu\nu} h, \label{metric_decom}
\end{align}
where the suffix $TT$ of a tensor stands for the transverse-traceless and the suffix $T$ of a vector stands for the transverse, that is,
\begin{align}
\partial^{\mu}h^{TT}_{\mu\nu}=0,~h^{TT\mu}{}_{\mu}=0,~\partial^{\mu}h^T_{\mu}=0.
\end{align}
The torsion can be also decomposed into
\begin{align}
T_{\mu}&=A^T_{\mu}+\partial_{\mu}\phi , \\
\Tcal_{\mu}&=\Acal^T_{\mu}+\partial_{\mu}\varphi, \label{Tcal_decom}  \\
\Ta_{\mu\nu\rho}&=2\partial_{[\nu}t^{TT}_{\rho]\mu}+2\left(\frac{\partial_{\mu}\partial_{[\nu}}{\Box}-\frac{1}{3}\eta_{\mu\nu} \right) B^T_{\rho]}
\nn
&+\epsilon_{\nu\rho}{}^{\alpha\beta}\left[ \partial_{\alpha}\tau^{TT}_{\beta\mu}+\left(\frac{\partial_{\beta}\partial_{\mu}}{\Box}-\frac{1}{3}\eta_{\beta\mu} \right)\Bcal^T_{\alpha}\right]. \label{T1_decom}
\end{align}
These decompositions classify the metric perturbations and the torsion with respect to their spins and parities. 

\begin{widetext}
The quadratic action around the Minkowski background is then given by
\begin{align}
\mathcal{L}_G^{(2)}=\mathcal{L}_{2^+}+\mathcal{L}_{2^-}+\mathcal{L}_{1^+}+\mathcal{L}_{1^-}+\mathcal{L}_{0^+}+\mathcal{L}_{0^-},
\end{align}
where
\begin{align}
\mathcal{L}_{2^+}&=\frac{1}{2}(h^{TT}_{\alpha\beta},t^{TT}_{\alpha\beta})
\begin{pmatrix}
\frac{1}{4}M_{\rm pl}^2+\frac{1}{2}\alpha_2 \Box & \frac{1}{2}\beta_1 \Box \\
* & -2a_1 M_T^2+2(2b_1+b_3)\Box
\end{pmatrix}
\Box
\begin{pmatrix}
h^{TT\alpha\beta} \\
t^{TT\alpha\beta} 
\end{pmatrix}
,\\
\mathcal{L}_{2^-}&=\frac{1}{2}\tau^{TT}_{\alpha\beta}(2a_1M_T^2-4b_1 \Box)\Box \tau^{TT\alpha\beta}
, \\
\mathcal{L}_{1^+}&=\frac{1}{2}(\Acal^T_{\alpha},\Bcal^T_{\alpha})
\begin{pmatrix}
a_3 M_T^2 -2b_6 \Box & -\frac{4}{3}b_9 \Box \\
* & -\frac{4}{3}a_1 M_T^2 +\frac{4}{9}(6b_1+4b_2+b_3)\Box 
\end{pmatrix}
\begin{pmatrix}
\Acal^{T\alpha} \\
\Bcal^{T\alpha}
\end{pmatrix}
, \\
\mathcal{L}_{1^-}&=\frac{1}{2}(A^T_{\alpha},B^T_{\alpha})
\begin{pmatrix}
a_2M_T^2-2b_4 \Box & -\frac{2}{3}b_8 \Box \\
* & \frac{4}{3}a_1 M_T^2 -\frac{8}{9}(3b_1+2b_2+b_3)\Box 
\end{pmatrix}
\begin{pmatrix}
A^{T\alpha} \\
B^{T\alpha}
\end{pmatrix}
,\\
\mathcal{L}_{0^+}&=\frac{1}{2}(\Phi,\phi)
\begin{pmatrix}
-\frac{1}{2}M_{\rm pl}^2 +2(3\alpha_1+\alpha_2)\Box & -\sqrt{3} \beta_2 \Box \\
* &  -a_2 M_T^2+2(b_4+b_5)\Box
\end{pmatrix}
\Box
\begin{pmatrix}
\Phi \\
\phi
\end{pmatrix}
,\\
\mathcal{L}_{0^-}&=\frac{1}{2}\varphi\left( -a_3 M_T^2+2(b_6+b_7)\Box \right)\Box \varphi
,
\end{align}
with
\begin{align}
\Phi=\frac{4}{\sqrt{3}}(h-\Box \sigma)=\frac{1}{\sqrt{3}}\theta_{\mu\nu}h^{\mu\nu}
\end{align}
and $\theta_{\mu\nu}:=\eta_{\mu\nu}-\frac{\partial_{\mu}\partial_{\nu}}{\Box}$. The symbol $*$ is attached to omit the symmetric parts of the matrices. In the momentum space, we define following quantities
\begin{align}
K_{2^+}&:=
-q^2 
\begin{pmatrix}
\frac{1}{4}M_{\rm pl}^2-\frac{1}{2}\alpha_2 q^2 & -\frac{1}{2}\beta_1 q^2 \\
* & -2a_1 M_T^2-2(2b_1+b_3)q^2
\end{pmatrix}
, \\
K_{2^-}&:=-q^2(2a_1 M_T^2 + 4b_1 q^2)
, \\
K_{1^+}&:=
\begin{pmatrix}
a_3 M_T^2 +2b_6 q^2 & \frac{4}{3}b_9 q^2 \\
* & -\frac{4}{3}a_1 M_T^2 -\frac{4}{9}(6b_1+4b_2+b_3)q^2 
\end{pmatrix}
, \\
K_{1^-}&:=
\begin{pmatrix}
a_2M_T^2+2b_4 q^2 & \frac{2}{3}b_8 q^2 \\
* & \frac{4}{3}a_1 M_T^2 +\frac{8}{9}(3b_1+2b_2+b_3)q^2 
\end{pmatrix}
, \\
K_{0^+}&:=
-q^2
\begin{pmatrix}
-\frac{1}{2}M_{\rm pl}^2 -2(3\alpha_1+\alpha_2)q^2 & \sqrt{3}\beta_2 q^2 \\
* &  -a_2 M_T^2-2(b_4+b_5)q^2
\end{pmatrix}
,\\
K_{0^-}&:=-q^2(-a_3 M_T^2-2(b_6+b_7)q^2).
\end{align}
\end{widetext}
Then, the number of poles of $K^{-1}_a~(a=0^{\pm},1^{\pm},2^{\pm})$ basically correspond to the number of particle species in each sector although there are unphysical poles at $q=0$ in $K_{2^-}^{-1},K_{0^-}^{-1},K_{0^+}^{-1}$. The unphysical poles of $K_{2^-}^{-1},K_{0^-}^{-1}$ are due to the presence of the derivative in the expressions \eqref{Tcal_decom} and \eqref{T1_decom} while that of $K_{0^+}^{-1}$ is due to not only the derivatives but also the remaining gauge mode. A more rigorous counting of the number of particle species will be performed based on the $3+1$ decomposition in the next section. For generic choices of parameters, the number of particle species in each sector is as follows: 
\begin{align*}
2^+&:2+1~\mbox{particle species}~(2~{\rm massive}+1~{\rm massless}), 
\\
2^-&:1~\mbox{particle species}~({\rm massive}),
\\
1^+&:2~\mbox{particle species}~({\rm massive}),
\\
1^-&:2~\mbox{particle species}~({\rm massive}),
\\
0^+&:2~\mbox{particle species}~({\rm massive}),
\\
0^-&:1~\mbox{particle species}~({\rm massive}).
\end{align*}
Here, $s^{\pm}$ on the left denotes the spin $s$ and the parity $\pm$. In any Lorentz-invariant theories the number of local physical degrees of freedom for each particle species for $s=(2, 1, 0)$ is $N^{m\ne 0}_s=(5, 3, 1)$ for massive cases, and $N^{m=0}_s=(2, 2, 1)$ for massless cases. Therefore the total number of local physical degrees of freedom is $(2\times 5 + 1\times 2) + (1\times 5) + (2\times 3) + (2\times 3) + (2\times 1) + (1\times 1) = 32$. For the massive cases the masses of these particles can be evaluated by the locations of the poles.

When the coupling constants satisfy a critical condition, a pole vanishes correspondingly to a reduction of the number of degrees of freedom\footnote{The limit to satisfy a critical condition, say $b_1 \rightarrow 0$, leads to the infinite mass of the corresponding particle specie. Therefore, in the sense of EFT, the critical conditions are not necessary to hold exactly for the reduction of the number of particle species. It suffices to satisfy the critical condition approximately, as far as the mass of the corresponding particle species is sufficiently heavier than the energy scale of interest.}. In the $2^+,1^{\pm},0^+$ sectors, after assuming the critical condition, the degrees of freedom can be further reduced by imposing an additional critical condition. We shall call these critical conditions the primary critical condition and the secondary critical condition in order. All critical conditions are summarized in Table \ref{Table1}.

In the case of PGT, the coupling constants satisfy all primary critical conditions and then the number of particle species in each sector is 
\begin{align*}
2^+&:1+1~\mbox{particle species}~(1~{\rm massive}+1~{\rm massless}), 
\\
2^-&:1~\mbox{particle species}~({\rm massive}),
\\
1^+&:1~\mbox{particle species}~({\rm massive}),
\\
1^-&:1~\mbox{particle species}~({\rm massive}),
\\
0^+&:1~\mbox{particle species}~({\rm massive}),
\\
0^-&:1~\mbox{particle species}~({\rm massive}),
\end{align*}
when no additional assumptions are imposed. Other critical conditions in PGT are summarized in Table \ref{Table2}. In the limit $M_T\rightarrow \infty$, the locations of some poles go to infinity and thus the dynamical ones in this limit are
\begin{align*}
2^+&:1+1~\mbox{particle species}~(1~{\rm massive}+1~{\rm massless}), 
\\
0^+&:1~\mbox{particle species}~({\rm massive}),
\end{align*}
and there is no particle species in the $2^-$, $1^{\pm}$ and $0^-$ sectors. This is, needless to say, consistent with the well-known content of particle species for the Riemannian quadratic curvature theory \eqref{quadratic_gravity}.

\begin{table}[htb]
\centering
\caption{Critical conditions in general theory.}
\label{Table1}
  \begin{tabular}{cccc} \hline \hline
               & \multicolumn{2}{c}{Critical conditions} \\ \hline
    $2^+$      & $4\alpha_2(2b_1+b_3)-\beta_1^2=0$ & (primary) \\
               & $(2b_1+b_3)M_{\rm pl}^2-2a_1 \alpha_2 M_T^2=0$ &(secondary) \\ 
    $2^-$      & $b_1=0$ &                                              \\ 
    $1^+$      & $b_6(6b_1+4b_2+b_3)+2b_9^2=0 $ &(primary) \\
               & $ a_3(6b_1+4b_2+b_3)+6a_1 b_6 =0$ & (secondary) \\
    $1^-$      & $4b_4(3b_1+2b_2+b_3)-b_8^2=0$ &(primary) \\
               & $a_2 (3b_1+2b_2+b_3)+3a_1 b_4=0$ &(secondary) \\ 
    $0^+$      & $4(3\alpha_1+\alpha_2)(b_4+b_5)-3\beta_2^2=0 $ &(primary) \\
               & $  (b_4+b_5)M_{\rm pl}^2+2a_2  (3\alpha_1+ \alpha_2)M_T^2 =0$ & (secondary) \\ 
    $0^-$      & $b_6+b_7=0$ & \\ \hline
  \end{tabular}
\end{table}

\begin{table}[htb]
\centering
\caption{Critical conditions in Poincar\'{e} gauge theory}
\label{Table2}
  \begin{tabular}{cccc} \hline \hline
               & Critical conditions \\ \hline 
    $2^+$      & $(2r_1-2r_3+r_4)(t_1+\lambda)=0$ \\ 
    $2^-$      & $r_1=0$ \\ 
    $1^+$      & $(2r_3+r_5)(t_1+t_2)=0$ \\
    $1^-$      & $(r_1+r_4+r_5)(t_1+t_3)=0$ \\  
    $0^+$      & $(r_1-r_3+2r_4)(t_3-\lambda)=0$ \\ 
    $0^-$      & $r_2=0$ \\ \hline
  \end{tabular}
\end{table}

\subsection{Matter scattering via gravitational interaction}

By the use of the tetrad $e^a_{\mu}=\delta^a_{\mu}+\delta e^a_{\mu}$ and the spin connection $\omega^{ab}{}_{\mu}$, the matter coupling to gravity is given by
\begin{align}
\mathcal{L}_{\rm int}=\delta e^a{}_{\mu} \Theta^{\mu}{}_a - \frac{1}{2}\omega^{ab}{}_{\mu} S^{\mu}{}_{ab}
\end{align}
where $\Theta^{\mu}{}_a$ and $S^{\mu}{}_{ab}$ are the canonical energy-momentum tensor and the canonical spin tensor defined by
\begin{align}
\Theta^{\mu}{}_a:= \frac{1}{{\rm det}\, e} \frac{\delta S_{\rm m}}{\delta e^a{}_{\mu}}, \quad
S^{\mu}{}_{ab}:= -\frac{2}{{\rm det}\, e} \frac{\delta S_{\rm m}}{\delta \omega^{ab}{}_{\mu}}.
\end{align}
The canonical spin tensor has the antisymmetric indices $S^{\mu}{}_{ab}=-S^{\mu}{}_{ba}$ while the canonical energy-momentum tensor is not symmetric, in general. Around the Minkowski background, one does not need to distinguish the spacetime indices, $\alpha,\beta,\mu,\nu,\cdots,$ and the Lorentz indices, $a,b,\cdots,$ since they are just transformed by $\delta^a_{\alpha}$. The energy, momentum and angular momentum of the matter are supposed to be conserved on shell:
\begin{align}
\partial_{\mu}\Theta^{\mu}{}_{\nu}=0, \quad \partial_{\mu}S^{\mu}{}_{\alpha\beta}+\Theta_{\alpha\beta}-\Theta_{\beta\alpha}=0.
\end{align}
On the other hand, after changing the variables $(\delta e^a_{\mu},\omega^{ab}{}_{\mu}) \rightarrow ( \delta g_{\mu\nu},T_{\mu\nu\rho})$, we find
\begin{align}
\mathcal{L}_{\rm int}=\frac{1}{2}\delta g_{\mu\nu}T^{\mu\nu}-\frac{1}{4}T_{\mu\nu\rho}(S^{\mu\nu\rho}-S^{\nu\rho\mu}+S^{\rho\nu\mu})
\end{align}
where $T^{\mu\nu}$ is the Belinfante tensor,
\begin{align}
T^{\mu\nu}=\Theta^{\mu\nu}+\frac{1}{2}\partial_{\rho}(S^{\rho\mu\nu}-S^{\mu\rho\nu}+S^{\nu\mu\rho}),
\end{align}
which is symmetric and conserved
\begin{align}
T^{\mu\nu}=T^{\nu\mu}, \quad \partial_{\mu}T^{\mu\nu}=0.
\end{align}

Due to the conservation law, only the sectors $2^+$ and $0^+$ are mediators of gravity via the coupling $\delta g_{\mu\nu}T^{\mu\nu}$ at tree level. A matter particle with a non-vanishing canonical spin tensor is scattered by exchanging not only the graviton but also the torsion, called tordion, via the coupling to the spin tensor. As an example, let us consider the Dirac field 
\begin{align}
\mathcal{L}_D=\frac{i}{2} \left( \bar{\psi} \gamma^{\mu} \nablaA{}_{\mu} \psi - (\nablaA{}_{\mu} \bar{\psi}) \gamma^{\mu} \psi 
\right)-m\bar{\psi}\psi ,
\end{align}
where $\gamma_a$ is the gamma matrix with $\{ \gamma_a ,\gamma_b \}=-2\eta_{ab}$. The covariant derivative of a Dirac spinor is
\begin{align}
\nablaA{}_{\mu} \psi = \left( \partial_{\mu} + \frac{1}{8}\omega^{ab}{}_{\mu} [\gamma_{a}, \gamma_b ] \right) \psi
 \,. \label{D_psi}
\end{align}
The canonical spin tensor of a Dirac field is given by
\begin{align}
S^{\mu\nu\rho}=\frac{1}{2}\epsilon^{\mu\nu\rho\sigma}j^5_{\sigma} ,
\end{align}
where $j^5_{\mu}=\bar{\psi}\gamma_{\mu} \gamma^5 \psi$ is the axial current which is not conserved in general. As a result, the gravitational coupling is 
\begin{align}
\mathcal{L}_{\rm int}=\frac{1}{2}\delta g_{\mu\nu}T^{\mu\nu} - \frac{3}{4}\mathcal{T}^{\mu} j^5_{\mu},
\end{align}
and then the particle species with $1^+$ and $0^-$ in the gravity sector are mediators of gravity as well.

For simplicity, we consider a matter field whose canonical spin tensor vanishes, e.g., a minimal scalar field. The tree level scattering amplitude is then
\begin{align}
\mathcal{M}=\tilde{T}^{\mu\nu}(p_1,p_2) \Delta^{hh}_{\mu\nu,\rho\sigma}(q) \tilde{T}^{\rho\sigma}(p_3,p_4),
\end{align}
where $\tilde{T}^{\mu\nu}$ is the Fourier transform of the source with the external momenta $p_i~(i=1,2,3,4)$ and $q$ is the internal momentum. The gauge independent part of the graviton propagator is
\begin{align}
-i\Delta^{hh}_{\mu\nu,\rho\sigma} =i( K_{2^+}^{-1})^{11} P^{(2)}_{\mu\nu,\rho\sigma}+i(K_{0^+}^{-1})^{11}P^{(0)}_{\mu\nu,\rho\sigma},
\end{align}
where $( K_{2^+}^{-1})^{11}$ and $(K_{0^+}^{-1})^{11}$ are the upper left component of $K_{2^+}^{-1},K_{0^+}^{-1}$. The projection operators $P^{(2)}_{\mu\nu,\rho\sigma},P^{(0)}_{\mu\nu,\rho\sigma}$ are defined by
\begin{align}
P^{(2)}_{\mu\nu,\rho\sigma}&:=\theta_{\mu(\rho}\theta_{\sigma)\nu}-\frac{1}{3}\theta_{\mu\nu}\theta_{\rho\sigma}, \\
P^{(0)}_{\mu\nu,\rho\sigma}&:=\frac{1}{3}\theta_{\mu\nu}\theta_{\rho\sigma}.
\end{align}
In general, the propagator has a $q^{-4}$ behavior in the high energy limit,
\begin{align}
(K_{2^+}^{-1})^{11} &\rightarrow 8\left(4\alpha_2-\frac{\beta_1^2}{2b_1+b_3}\right)^{-1} \frac{1}{q^4}+O(q^{-6}),
\\
( K_{0^+}^{-1})^{11} &\rightarrow 2\left(4(3\alpha_1+\alpha_2)-\frac{3\beta_2^2}{b_4+b_5} \right)^{-1} \frac{1}{q^4} + O(q^{-6}),
\end{align}
which suggests there would be Ostrogradsky ghosts. On the other hand, when the primary critical conditions are imposed
\begin{align}
4\alpha_2(2b_1+b_3)-\beta_1^2=0,~ 4(3\alpha_1+\alpha_2)(b_4+b_5)-3\beta_2^2=0, \label{ghost_free}
\end{align}
the Ostrogradsky ghost modes may be eliminated and then the propagator recovers a $q^2$ behavior\footnote{Precisely, the second condition of \eqref{ghost_free} is not required to be free from the Ostrogradsky ghost. There is a massless spin-0 ghost but this massless ghost is harmless since this is just related to the remaining gauge mode as mentioned before (see \cite{Alvarez-Gaume:2015rwa} for example). It will, however, turn out in the next section that the second condition is indeed required in order that all dynamical modes are ghost-free. We therefore assume the second one as well.}. We recall that these conditions are automatically satisfied in PGTs. This could be understood by the fact that PGTs do not have the second derivatives of the tetrad in the Lagrangian (although this property is certainly spoiled by radiative corrections), and thus they are not higher derivative theories. The graviton propagator is then
\begin{align}
-i\Delta^{hh}_{\mu\nu,\rho\sigma} &= \frac{4}{M_{\rm pl}^2} \mathcal{D}_{\mu\nu,\rho\sigma}
+\frac{8m_{2^+}^2\alpha_2 }{M_{\rm pl}^4 } \frac{-i}{q^2+m_{2^+}^2} P^{(2)}_{\mu\nu,\rho\sigma}
\nn
&+\frac{8m_{0^-}^2(3\alpha_1+\alpha_2)}{M_{\rm pl}^4} \frac{-i}{q^2+m_{0^+}^2} P^{(0)}_{\mu\nu,\rho\sigma},
\label{propagator}
\end{align}
where $\mathcal{D}_{\mu\nu,\rho\sigma}$ is the graviton propagator of the Einstein gravity,
\begin{align}
\mathcal{D}_{\mu\nu,\rho\sigma}=\frac{-i}{q^2}\left( \theta_{\mu(\rho}\theta_{\sigma)\nu}-\frac{1}{2}\theta_{\mu\nu}\theta_{\rho\sigma} \right).
\end{align}
The masses of the massive spin-2 mode and of the massive spin-0 mode are
\begin{align}
m_{2^+}^2&=\frac{4a_1 \alpha_2 M_{\rm pl}^2 M_T^2}{M_{\rm pl}^2 \beta_1^2-8a_1 \alpha_2^2 M_T^2}
, \nn
m_{0^+}^2&=\frac{2a_2(3\alpha_1+\alpha_2)M_{\rm pl}^2 M_T^2}{8a_2 (3\alpha_1+\alpha_2)^2 M_T^2+3 \beta_2^2 M_{\rm pl}^2 } .
\end{align}

In the limit $M_T \gg M_{\rm pl}$, the masses become $m_{2^+}^2 \rightarrow -M_{\rm pl}^2/2\alpha_2$ and $m_{0^+}^2 \rightarrow M_{\rm pl}^2/4(3\alpha_1+\alpha_2)$, and then the propagator becomes
\begin{align}
\left. -i \Delta^{hh}_{\mu\nu,\rho\sigma} \right|_{M_T\rightarrow \infty} 
&= \frac{4}{M_{\rm pl}^2} \mathcal{D}_{\mu\nu,\rho\sigma}
-\frac{4}{M_{\rm pl}^2 } \frac{-i}{q^2+m_{2^+}^2} P^{(2)}_{\mu\nu,\rho\sigma}
\nn
&+\frac{2}{M_{\rm pl}^2} \frac{-i}{q^2+m_{0^+}^2} P^{(0)}_{\mu\nu,\rho\sigma}. \label{QG_prop}
\end{align}
This is indeed the well-known graviton propagator of the quadratic curvature theory \eqref{quadratic_gravity} with a massive spin-2 ghost. We note that the spin-2 part of \eqref{QG_prop} has a $q^{-4}$ behavior in the high energy limit which makes the theory renormalizable although the corresponding energy scale goes beyond the mass of the ghost mode~\cite{Stelle:1976gc}.

One may recover the $q^{-4}$ behavior if one removes the critical conditions \eqref{ghost_free}. However, in this case the Ostrogradsky ghosts appear below the scale at which the $q^{-4}$ behavior is realized. Therefore, in the present paper, we shall not discuss the renormalizability of graviton loops any more. Instead, we simply interpret the scale at which a ghost appears as the cutoff scale of the theory\footnote{Alternatively, there are attempts to obtain the Lorentzian metric signature from a Euclidean metric as a macroscopic effective description~\cite{Mukohyama:2013ew,Mukohyama:2013gra}. In this case, higher derivatives do not necessarily lead to the ghost problem. Another attempt is a modification of a quantization prescription to turn a ghost particle into a fake particle \cite{Anselmi:2017ygm}.}.

\section{Stability conditions}
\label{sec_stability}
The propagator \eqref{propagator} implies that the theory can be ghost-free if the inequalities $m_{2^+}^2>0,m_{0^+}^2>0$ and $\alpha_2>0,3\alpha_1+\alpha_2>0$ are satisfied. However, gravity is also mediated by particles of other sectors when the canonical spin tensor does not vanish. We thus investigate ghost-free conditions for all sectors of the Lagrangian \eqref{LG} under the assumptions 
\begin{align}
\alpha_1 \neq 0, \quad \alpha_2 \neq 0 .
\end{align}
We have already set $\alpha_3=0$ by using the Gauss-Bonnet theorem.

As discussed in the previous section, there still exists a remaining gauge mode when the metric perturbations are decomposed as \eqref{metric_decom}. We thus give up the explicit Lorentz invariance here; instead, we adopt the $3+1$ decompositions and then decompose perturbations into scalar-vector-tensor (SVT) components commonly used in the context of the cosmological perturbation theory. Although we discuss the perturbations around the Minkowski background, the $3+1$ decomposition can be straightforwardly extended to an arbitrary spacetime (see Appendix~\ref{append_3+1}).

We denote the $3+1$ components of the torsion as follows:
\begin{align}
T_0&=\phi, \quad T_i=A_i 
,
\\
\Tcal_0&=\varphi ,  \quad \Tcal_i = \Acal_i
,
\\
\Ta_{00i}&=B_i , \quad \Ta_{0ij}=\epsilon_{ijk}\Bcal^k
,
\nn
\Ta_{ij0}&=t_{ij}-\frac{1}{2}\epsilon_{ijk}\Bcal^k ,
\quad
\Ta_{ijk}=\delta_{i[j}B_{k]}+\epsilon_{jk}{}^{l}\tau_{il},
\end{align}
where $t_{ij}$ and $\tau_{ij}$ are symmetric and traceless. The metric perturbations $\delta g_{\mu\nu}=g_{\mu\nu}-\eta_{\mu\nu}$ are
\begin{align}
\delta g_{00}=-2\alpha, \quad \delta g_{0i}=\beta_i , \quad \delta g_{ij}=2(h_{ij}+\zeta \delta_{ij}),
\end{align}
where $h^i{}_i=\delta^{ij}h_{ij}=0$. 

When $\alpha_2 \neq 0$, the quadratic order Lagrangian $\mathcal{L}^{(2)}$ contains the second time derivative of $h_{ij}$. It is then useful to eliminate second time derivatives and to reduce the action to the form that depends on perturbation variables and their first derivatives only by introducing auxiliary fields and by writing the action as
\begin{align}
\mathcal{L}'=\mathcal{L}^{(2)}|_{\ddot{h}_{ij}=\Phi_{ij}}+H_{ij}(\Phi^{ij}-\ddot{h}^{ij}),
\end{align}
where $H_{ij},\Phi_{ij}$ are symmetric and traceless. The variation with respect to $H_{ij}$ yields the constraint $\Phi_{ij}=\ddot{h}_{ij}$ and thus the new Lagrangian is equivalent to the original one. Instead, one can take the integration by part so that
\begin{align}
\mathcal{L}'=\mathcal{L}^{(2)}|_{\ddot{h}_{ij}=\Phi_{ij}}+H_{ij}\Phi^{ij}+\dot{H}_{ij}\dot{h}^{ij} + (\mbox{total derivative}), \label{L_with_H}
\end{align}
and can take the variation with respect to $\Phi^{ij}$. One then obtains the constraint which determines $\Phi^{ij}$ in terms of other variables. After substituting the constraint into the Lagrangian, one finds the Lagrangian with up to the first order time derivative of $h_{ij}$.

In the SVT decompositions, a scalar, a vector and a traceless tensor are decomposed into scalar type $(S)$, vector type $(V)$, and tensor type $(T)$ perturbations in terms of the spatial rotation:
\begin{align}
\phi&=\phi^{(S)},\quad A_i=\partial_i A^{(S)}+ A^{(V)}_i, \nn
t_{ij}&=\left( \partial_i \partial_j -\frac{1}{3}\delta_{ij} \partial^2 \right) t^{(S)}+\partial_{(i} t^{(V)}_{j)}+t^{(T)}_{ij},
\end{align}
where 
\begin{align}
\partial^i A^{(V)}_i=\partial^i t^{(V)}_i=0, \quad \partial^i t^{(T)}_{ij}=0,\quad t^{(T)i}{}_{i}=0,
\end{align}
and SVT perturbations are decoupled from each other at the linear order in perturbations. The spatial dependence of the variables can be Fourier transformed and then each Fourier sector is characterized by the spatial momentum $k^i$. For each $k^i$, the vector type and the tensor type perturbations have two different modes which can be chosen as the helicity $\pm 1 $ modes and the helicity $\pm 2$ modes, respectively. Let us denote two helicity basis $Y_R^i(k),Y_L^i(k)$ for vectors and $Y_R^{ij}(k),Y_L^{ij}(k)$ for tensors, respectively. The vector perturbations ($A^{(V)}_i$, $B^{(V)}_i$, $t^{(V)}_i$) and the pseudovector perturbations ($\Acal^{(V)}_i$, $\Bcal^{(V)}_i$, $\tau^{(V)}_i$) can be decomposed as
\begin{align}
 A^{(V)i}&=A_L Y_L^i + A_R Y_R^i, \nn
 B^{(V)i}&=B_L Y_L^i + B_R Y_R^i, \nn
 t^{(V)i}&=t^{(V)}_L Y_L^i + t^{(V)}_R Y_R^i, \\
\Acal^{(V)i}&=\Acal_L Y_L^i - \Acal_R Y_R^i, \nn
 \Bcal^{(V)i}&=\Bcal_L Y_L^i - \Bcal_R Y_R^i, \nn
 \tau^{(V)i}&=\tau^{(V)}_L Y_L^i - \tau^{(V)}_R Y_R^i.  
\end{align}
Similarly, tensor perturbations $t^{(T)}_{ij}$ and pseudotensor perturbations $\tau^{(T)}_{ij}$ are decomposed as
\begin{align}
t^{(T)ij}&= t_L Y_L^{ij}+t_R Y_R^{ij} ,\\ 
\tau^{(T)ij}&=\tau_L Y_L^{ij}-\tau_R Y_R^{ij}.
\end{align}
Note that there are couplings such as
\begin{align}
\epsilon^{ijk}A_i \partial_j \Acal_k, \quad
\epsilon^{ijk}t_{il}\partial_j \tau_k{}^{l},
\end{align}
due to the presence of the pseudovectors and the pseudotensors which give mixings between vectors and pseudovectors and those between tensors and pseudotensors, e.g. couplings between $A_L$ and $\Acal_L$. Needless to say, the L sector and the R sector are still decoupled from each other. Also, the L sector and the R sector obey the same equations of motion because of the absence of parity violating operators. In order to make this fact manifest, the minus sign has been inserted in front of the R modes in the above decomposition of the pseudovector and pseudotensor perturbations so that
\begin{align}
-\frac{1}{k}\epsilon^{ijk}\partial_j \Acal^{(V)}_k&=\Acal_L Y_L^i+\Acal_R Y_R^i
, \\
-\frac{1}{k}\epsilon^{(i|jk}\partial_j \tau^{(T)}{}_k{}^{l)}&=\tau_L Y_L^{il} +\tau_R Y_R^{il}
,
\end{align}
and so on.
 Contrary to the vector and tensor perturbations, the scalar perturbations have only one helicity mode. The parity preservation then guarantees that the (true) scalar perturbations and the pseudoscalar perturbations are decoupled. As a result, the general perturbations can be classified into four decoupled sectors, namely the tensor perturbations (which include both the truetensor and pseudotensor perturbations), the vector perturbations (which include both truevector and pseudovector perturbations), the (true) scalar perturbations, and the pseudoscalar perturbations. In Appendix~\ref{append_ghost}, we summarize how to confirm absence or existence of a ghost instability and how to evaluate masses of the particles from the quadratic Lagrangian of each sector.

\subsection{Tensor perturbations}
As discussed in Section \ref{sec_Lorentz_inv}, there generally exists the Ostrogradsky ghost due to the presence of higher time derivatives. However, this can be removed by imposing the critical condition
\begin{align}
4\alpha_2 (2b_1+b_3) - \beta_1^2=0 . \label{deg1}
\end{align}
This kind of conditions is called degeneracy conditions in the context of higher order scalar-tensor theories~\cite{Langlois:2015cwa} since these are conditions under which the kinetic matrix is degenerate. In Appendix~\ref{append_toy}, we study a toy model that explains the reason why higher time derivatives lead to the Ostrogradsky ghost and the reason why the ghost can be removed by the degeneracy condition. In this sense, what we seek here is a degenerate higher order spin-2 theory.

After integrating out $\Phi_{ij}$ from \eqref{L_with_H}, we confirm that there exists the Ostrogradsky ghost when \eqref{deg1} is not imposed. We then impose \eqref{deg1} under which the tensor perturbations of $t_{ij}$ are non-dynamical. After integrated out $t_{ij}$, the quadratic Lagrangian of the tensor perturbations becomes
\begin{align}
\mathcal{L}_T&=\frac{M_{\rm pl}^2}{2}\left[ \dot{\tilde{h}}_{ij}^2 -k^2 \tilde{h}_{ij}^2 \right]
\nn
&+\left(\frac{(2b_1+b_3)^2}{a_1\beta_1^2 M_T^2}-\frac{1}{2M_{\rm pl}^2}  \right) \left[ \dot{H}_{ij}^2-(k^2+m_{2^+}^2) H_{ij}^2 \right]
\nn
&-\frac{2a_1 b_1 M_T^2 }{2b_1k^2+a_1 M_T^2}\left[ \dot{\tilde{\tau}}_{ij}^2-\left(k^2+\frac{a_1}{2b_1} M_T^2\right) \tilde{\tau}_{ij}^2 \right],
\end{align}
with new variables
\begin{align}
\tilde{h}_{ij}&=h_{ij}+\frac{1}{M_{\rm pl}^2}H_{ij},
\\
\tilde{\tau}_{ij}&=\tau_{ij} -\frac{2b_1+b_3}{a_1 \beta_1 M_T^2} \epsilon_{(i|kl}\partial^k H_{j)}{}^l ,
\end{align}
where the suffix $(T)$ has been dropped for simplicity of notation. In the limit $k^2 \gg a_1 M_T^2 /b_1$, $H_{ij}$ or $\tilde{\tau}_{ij}$ is a ghost depending on whether $a_1<0$ or $a_1>0$. In the case $a_1<0$, one cannot assume the critical condition of $2^+$ to remove the ghost $H_{ij}$. Therefore, we have to assume $a_1>0$ and the critical condition of the $2^-$ sector,
\begin{align}
b_1=0 . \label{deg2}
\end{align}
Regardless of whether the massive $2^+$ mode is dynamical or not, the $2^-$ sector must be non-dynamical in order to be ghost-free if $\alpha_2 \neq 0$. 

As a result, the stability conditions of the tensor perturbations require the conditions \eqref{deg1} and \eqref{deg2}, namely the primary critical condition of $2^+$ and the critical condition of $2^-$. The ghost-free and tachyon-free conditions are then
\begin{align}
\alpha_2>0,\quad 0<a_1 M_T^2<\frac{  \beta_1^2}{8 \alpha_2^2}M_{\rm pl}^2.
\end{align}

\subsection{Scalar and pseudoscalar perturbations}
Analysis of vector perturbations is usually easier than analysis of scalar ones because the number of independent multiplets of the vector ones is usually less than that of the scalar ones. However, in the present case, perturbations of (true) scalars and pseudoscalars are decoupled due to the parity conservation while the perturbations of (true) vectors and pseudovectors are coupled. As a result, the analysis of the vector perturbations are more involved compared with the scalar and pseudoscalar ones. In practice, the easiest one among the three (vector, scalar and pseudoscalar) types of perturbations is the analysis of the pseudoscalar perturbations in the present case. We therefore first present the analysis of the pseudoscalar perturbations and then the scalar perturbations. After imposing the stability conditions for these two sectors, it is relatively easy to show the stability of the vector perturbations. Indeed, it will turn out that the stability of the vector perturbations does not require additional conditions, provided that the stability conditions for the pseudoscalar and scalar perturbations are imposed in advance.

\subsubsection{Pseudoscalar perturbations}
There are four variables from $\Acal_i,\Bcal_i,\varphi$ and $\tau_{ij}$ in the pseudoscalar perturbations. We have seen that the ghost-free condition of the tensor perturbations requires $b_1=0$ so that $\tau_{ij}$ is non-dynamical. After integrating out pseudoscalar perturbations of $\tau_{ij}$, we still find a ghost instability in the limit $k\rightarrow \infty$. We thus assume one of the conditions
\begin{align}
b_6(4b_2+b_3)+2b_{9}^2=0, \label{cp1+}
\end{align}
or
\begin{align}
b_6+b_7=0, \label{cp0-}
\end{align}
namely the primary critical condition of $1^+$ or the critical condition of $0^-$. In either case, the pseudoscalar perturbations can be stable in an appropriate parameter space. Let us call theories with \eqref{cp1+} Class I and theories with \eqref{cp0-} Class II, respectively. We recall that PGTs satisfy all primary critical conditions and thus they are automatically classified into Class I.

In Class I theories, the critical conditions of $2^-$ and $1^+$ have been assumed and thus the pseudoscalar perturbations must have the particles with $1^+$ and $0^-$ only. Indeed, we confirm that there are two dynamical variables with masses
\begin{align} 
m_{1^+}^2=\frac{3a_1 a_3}{a_3(4b_2+b_3)+6a_1 b_6}M_T^2, \quad
m_{0^-}^2=\frac{a_3}{2(b_6+b_7)}M_T^2,
\end{align}
whose values are consistent with those obtained by the positions of the poles of $K^{-1}_{1^+}$ and $K^{-1}_{0^-}$ under the critical conditions. The stability conditions of these modes are
\begin{align}
a_3>\frac{12 a_1 b_{9}^2}{(4b_2+b_3)^2},~ 4b_2+b_3>0,~ b_6+b_7>0.
\end{align}
Note that $a_3$ has to be positive since $a_1>0$ has been imposed from the stability condition of the tensor perturbations.

\begin{widetext}
On the other hand, the $0^-$ sector is non-dynamical while both particles in the $1^+$ sector are dynamical in Class II theories. In this case, $\mathcal{T}^{\mu}$ has the Maxwell-type kinetic term:
\begin{align}
b_6 (\nabla_{\mu}\Tcal_{\nu} \nabla^{\mu} \Tcal - \nabla_{\mu}\Tcal^{\mu} \nabla_{\nu} \Tcal^{\nu})=\frac{b_6}{2} \mathcal{F}_{\mu\nu}\mathcal{F}^{\mu\nu} + (\mbox{total derivative}),
\end{align}
where $\mathcal{F}_{\mu\nu}=2\nabla_{[\mu}\Tcal_{\nu]}$. The ghost and tachyon free conditions are
\begin{align}
a_3<0,~ (4b_2+b_3)b_6+2b_{9}^2<0,~4b_2+b_3>0.
\end{align}
The masses of these particles are given by the roots of 
\begin{align}
\{ 4b_{9}^2+2b_6(4b_2+b_3)\} m^4-\{ a_3(4b_2+b_3)+6a_1 b_6\} m^2 M_T^2+3a_1 a_3 M_T^4=0.
\end{align}
In particular, when $b_{9}=0$, $\Tcal_{\mu}$ is decoupled and the masses are
\begin{align}
m_{\mathcal{T}}^2=\frac{a_3}{2b_6}M_T^2,~ m_{1^-}^2=\frac{3a_1}{4b_2+b_3}M_T^2.
\end{align}
\end{widetext}

It is worth mentioning that both Class I and Class II theories have to satisfy the inequality
\begin{align}
4b_2+b_3>0
\end{align}
to be free from the ghost. Combining the ghost-free condition of the tensor mode $b_3>0$, one obtains an inequality
\begin{align}
2b_2+b_3>0
\end{align}
which will be imposed hereafter, especially for the analysis of stability conditions of the scalar perturbations.

\subsubsection{Scalar perturbations}
We then analyze (true) scalar perturbations which generally contain degrees of freedom from the $2^+,1^-,0^+$ sectors. Note that there are gauge degrees of freedom in scalar perturbations. We choose the gauge $\beta_i=0,\zeta=0$ so that the quadratic order Lagrangian \eqref{L_with_H} contains time derivatives up to first order\footnote{Even if choosing another gauge, say $\beta_i=0,h_{ij}=0$, the second time derivatives of $\zeta$ disappear after integrating out $\alpha$.}. After eliminating non-dynamical variables $(\alpha, t_{ij},\Phi_{ij})$ under the ghost-free conditions obtained from tensor perturbations, we find a quadratic Lagrangian in terms of the variables $(h_{ij},H_{ij},A_i,B_i,\phi)$ and confirm that there is still a ghost, in general. 

We do not assume the secondary critical condition of $2^+$ so that the massive spin-2 particle is dynamical. There are then two possibilities to get a degenerate kinetic matrix: imposing the critical condition of $1^-$
\begin{align}
4b_4(2b_2+b_3)-b_8^2=0, \label{cp1-}
\end{align}
or imposing the critical condition of $0^+$,
\begin{align}
4(3\alpha_1+\alpha_2)(b_4+b_5)-3\beta_2^2=0, \label{cp0+}
\end{align}
both of which lead to a degenerate kinetic matrix. Contrary to the pseudoscalar perturbations, it will turn out that both conditions \eqref{cp1-} and \eqref{cp0+} have to be simultaneously imposed to obtain a ghost-free theory. Furthermore, we have to also impose the secondary critical condition of the $1^-$ sector
\begin{align}
3a_1b_4+a_2(2b_2+b_3)=0. \label{cs1-}
\end{align}

For instance, let us assume \eqref{cp1-} is satisfied so that $\phi$ becomes non-dynamical. After integrating out $\phi$, we find that there still exists a ghost mode in the limit $k\rightarrow \infty$. Therefore, we further require the degeneracy of the kinetic matrix. There are again two possibilities to degenerate the kinetic matrix; one is, of course, the condition \eqref{cp1+} and the other is the secondary critical condition of $0^+$,
\begin{align}
(b_4+b_5)M_{\rm pl}^2+2a_2  (3\alpha_1+ \alpha_2)M_T^2 =0. \label{cs0+}
\end{align}
Even if \eqref{cs0+} is assumed, the ghost cannot be removed. Therefore, we have to assume \eqref{cp1-} in order to remove the ghost degree of freedom. On the other hand, when the condition \eqref{cp1-} is firstly assumed without imposing \eqref{cp0+}, there is a non-dynamical variable which is a linear combination of the variables $(H_{ij}, h_{ij}.A_i,B_i)$. The variation with respect to it yields a constraint equation which generally determines the non-dynamical variables in terms of other variables. However, after the non-dynamical variable is eliminated by using the constraint equation, the ghost cannot be removed. The only way to remove the ghost is to impose the conditions \eqref{cp0+} and \eqref{cs1-} simultaneously so that the non-dynamical variable becomes a Lagrangian multiplier\footnote{The linearity of the action with respect to a non-dynamical variable is indeed crucial for the removal of the Boulware-Deser ghost in massive spin-2 theories (see \cite{deRham:2014zqa,Schmidt-May:2015vnx} for reviews and references therein).}. 

Imposing three critical conditions \eqref{cp1-}, \eqref{cp0+} and \eqref{cs1-}, we find that there are two non-dynamical variables, one of which is a Lagrangian multiplier. Therefore, three of five variables $(h_{ij},H_{ij},A_i,B_i,\phi)$ can be eliminated by solving the constraint equations. We finally obtain the quadratic order Lagrangian which contains two dynamical degrees of freedom from the $2^+,0^+$ sectors. The stability conditions are, as expected, the inequalities $m_{2^+}^2>0,m_{0^+}^2>0,\alpha_2>0,3\alpha_1+\alpha_2>0$. Note that $a_1>0,2b_2+b_3>0$ have to be satisfied from the stability conditions of the tensor and pseudoscalar perturbations which then lead to $b_4>0,a_2<0$ via \eqref{cp1-} and \eqref{cs1-}. As a result, the inequality $m_{0^+}^2>0$ is guaranteed by
\begin{align}
 a_2 M_T^2 < -\frac{3}{8}\frac{\beta_2^2}{(3\alpha_1+\alpha_2)^2}M_{\rm pl}^2.
\end{align}

\subsection{Vector perturbations}
We assume the ghost and tachyon free conditions obtained from the tensor, scalar, and pseudoscalar perturbations. It is then relatively straightforward to confirm that there is no unstable modes in the vector perturbations in this case. We also confirm that Class I theory has two dynamical degrees of freedom in the vector perturbations while Class II theory has three dynamical degrees of freedom. This is consistent with the discussion of Sec.~\ref{sec_Lorentz_inv}: vector perturbations of Class I theory are originated from $2^+$ and $1^+$ while those of Class II are from $2^+$ and two of $1^+$.

\subsection{Summary of the stability conditions}
\label{subsec_summary}

Let us summarize the conditions for Class I and Class II theories and their particle contents. The common critical conditions are
\begin{align}
b_1&=0 ,\\ 
4\alpha_2 b_3 - \beta_1^2&=0 , \\
4(3\alpha_1+\alpha_2)(b_4+b_5)-3\beta_2^2&=0, \\
4b_4(2b_2+b_3)-b_8^2&=0 ,\\
3a_1b_4+a_2(2b_2+b_3)&=0, 
\end{align}
and both theories must satisfy the inequalities
\begin{align}
3\alpha_1+\alpha_2&>0
, \\
\alpha_2&>0
,\\
4b_2+b_3&>0
, 
\end{align}
and
\begin{align}
0<\frac{3}{-8a_2}\frac{\beta_2^2}{(3\alpha_1+\alpha_2)^2}M_{\rm pl}^2 < M_T^2 < \frac{1}{8a_1} \frac{  \beta_1^2}{\alpha_2^2}M_{\rm pl}^2,
\label{bound_MT}
\end{align}
in order that the Minkowski spacetime is stable. Class I theory is defined by the further critical condition
\begin{align}
b_6(4b_2+b_3)+2b_{9}^2=0,
\end{align}
with the stability conditions
\begin{align}
a_3>\frac{12 a_1 b_{9}^2}{(4b_2+b_3)^2},~ b_6+b_7>0.
\end{align}
On the other hand, Class II theory is obtained by imposing the critical condition,
\begin{align}
b_6+b_7=0,
\end{align}
and the stability conditions,
\begin{align}
a_3<0,~ (4b_2+b_3)b_6+2b_{9}^2<0.
\end{align}
As a result, Class I and Class II theories have the following massive degrees of freedom
\begin{align*}
2^+,1^+,0^+,0^- \quad &{\rm (Class~I) },
\\
2^+,1^+,1^+,0^+ \quad &{\rm (Class~II) },
\end{align*}
in addition to the massless spin-2 graviton. The particles with $2^+$ and $0^+$ are mediators of gravity via the coupling to the energy-momentum tensor while the $1^+$ and $0^-$ particles can be mediators when the spin tensor of matter fields does not vanish.

It would be worth noting that $M_T$ is bounded from above and below. The value of $M_T$ is not completely independent from the Planck mass. For simplicity, we assume $|\alpha_i| \sim \alpha, |\beta_i| \sim \beta, |b_i| \sim b$ where $\alpha,\beta,b$ represent the order of magnitude of each group of parameters. If all $a_i$ are the same order of the magnitude, the modulus of $a_i$ can be $O(1)$ by normalizing $M_T$. The ghost-free conditions then constrain $\alpha b \sim \beta^2$ and $M_T \sim \alpha^{-1} \beta M_{\rm pl}$. As a result, all massive particles have masses of the order of $\alpha^{-1/2} M_{\rm pl}$.

Most of the critical conditions are the primary ones. In particular, all primary critical conditions have to be imposed in Class I theory. Recall that the primary critical conditions are satisfied in PGTs and there are following relations
\begin{align}
b_1&=r_1
,\\
3b_1+2b_2+b_3&=r_1+r_4+r_5
, \\
\frac{3}{-8a_2}\frac{\beta_2^2}{(3\alpha_1+\alpha_2)^2}&=\frac{M_T^2}{4\lambda-2t_3}
, \\
\frac{1}{8a_1} \frac{  \beta_1^2}{\alpha_2^2}&=\frac{M_T^2}{4\lambda+2t_1}
.
\end{align}
The ghost-freeness of Class I theory requires the critical condition of $2^-$ and $1^-$ which are written as
\begin{align}
r_1=0,\quad (r_1+r_4+r_5)(t_1+t_3)=0,
\end{align}
in PGTs. Since $3b_1+2b_2+b_3=(r_1+r_4+r_5)>0$ must hold from the stability conditions, the second condition turns out to be
\begin{align}
t_1=-t_3.
\end{align}
This indicates that the ghost-free condition \eqref{bound_MT} cannot be satisfied in PGTs. In that case, one should further assume a critical condition to remove a ghost. Indeed, a ghost free theory can be found in PGTs when the critical condition of the $0^+$ sector,
\begin{align}
(r_1-r_3+2r_4)(t_3-\lambda)=0 ,
\end{align}
is imposed. When we assume the branch $t_3=\lambda$, the critical condition of $2^+$ is also satisfied and then the massive spin-2 mode becomes non-dynamical. On the other hand, the branch $r_1-r_3+2r_4=0$, which corresponds to $3\alpha_1+\alpha_2=\beta_2=0$, yields a ghost-free theory with a dynamical massive spin-2 mode. This model has been discussed in the literature~\cite{Hayashi:1979wj,Hayashi:1980ir,Hayashi:1980qp,Sezgin:1979zf,Nair:2008yh,Nikiforova:2009qr}.

This fact gives us an interesting observation. Both of the metric theory \eqref{quadratic_gravity} and PGTs \eqref{PGT} have the massive spin-2 particle and the massive spin-0 particle in their graviton propagator in general. The ghost-free condition of the metric theory requires $\alpha_2=0$ which kills the degree of freedom of the massive spin-2 mode of the propagator. On the other hand, in PGTs a ghost-free theory with the massive spin-2 particle can be obtained by imposing $3\alpha_1+\alpha_2=0$; this condition kills the massive spin-0 mode of \eqref{propagator}, instead. In either cases, all particles species that generically appear in the propagator \eqref{propagator} cannot be simultaneously non-ghost. However, in the extended theories introduced in the present paper, thanks to the new terms in the Lagrangian that depend on derivatives of the torsion, i.e. second derivatives of the tetrad, both massive spin-2 and massive spin-0 modes in \eqref{propagator} can coexist.

\section{Concluding remarks}
\label{summary}
The present paper has discussed derivative corrections to the Einstein gravity in the framework beyond the Riemannian geometry and found new classes of higher curvature theories with non-ghost massive spin-2 and spin-0 particle species around the Minkowski background. We firstly assume that the metric and the connection (or, the tetrad and the spin connection) are independent. In the particle physics sense, this extension of the spacetime geometry corresponds to the introduction of massive particles encoded in the torsion and the non-metricity tensors. If these massive particles are heavy enough to be integrated out then the Riemannian description of the spacetime is effectively recovered. Just for simplicity, we have considered theories with a metric compatible connection. We then take into account all possible terms in the Lagrangian up to scaling dimension four and identify the conditions under which ghosts and tachyons are avoided around the Minkowski background. The stability conditions and the particle contents of the ghost-free theories are summarized in \S.~\ref{subsec_summary}. A key feature of the new classes of theories is that both massive spin-2 and spin-0 particles can coexist with the massless spin-2 graviton without any instability. This cannot be achieved by either the metric theories or the Poincar\'{e} gauge theories (PGTs). As a result, the matter is gravitationally scattered by exchanging not only the massless spin-2 graviton but also the additional massive spin-2 and spin-0 particles sourced by the energy and momentum of the matter. The ghost-free theories are classified into two classes where the difference appears when a matter has a non-vanishing spin tensor like a Dirac field. Class I theory has additional massive spin-1 and spin-0 particle species which mediate the force sourced by the spin of matter fields. On the other hand, Class II theory has two spin-1 particle species which are also the mediators via the coupling to the spin. All massive particles would have the masses of the same order of the magnitude when no additional hierarchy of the parameters is assumed. For instance, the masses are of order of $10^{13}$GeV if the coefficients of the quadratic curvature terms are of the same order of the magnitude as the Starobinsky inflationary model.

Therefore, the result of the present paper potentially opens up a new window to phenomenological signatures of derivative corrections to the Einstein gravity within the regime of validity of an effective field theory (EFT). In particular, the torsionfull quadratic curvature theory predicts that a matter field with a non-vanishing spin tensor and one without the spin tensor feel different gravitational forces due to the $1^+$ and $0^-$ particle species that mediate gravitational forces via the spin tensor. This aspect of the no-ghost torsionfull theory is completely different from the torsionless theory. Therefore new types of experimental constraints may be obtained via the tests of the equivalence principle if the extra particle species in the gravitational sector are light enough.

Even if the extra particle species in the gravitational sector are relatively heavy, signatures of derivative corrections to the Einstein gravity in this framework may be found in the early universe. For example, it would be interesting to investigate implications of the extra massive particle species to the cosmological collider physics~\cite{Arkani-Hamed:2015bza} in the context of the Starobinsky inflationary model. However, for quantitative analysis, some extensions of the present result are required in order to apply it to cosmological scenarios. This is because we have only considered the linear perturbations around the Minkowski background and thus our results cannot be applied directly to other backgrounds, around which terms such as $\mathcal{L}_{RT^2},\mathcal{L}_{(\nabla T)T^2}$ and $\mathcal{L}_{T^4}$ may contribute to the quadratic action for perturbations. For example, to consider the inflationary universe, although $\mathcal{L}_{(\nabla T)T^2}$ and $\mathcal{L}_{T^4}$ could be negligible as long as a torsionless de Sitter background is considered, $\mathcal{L}_{RT^2}$ has to be taken into account to study quantum fluctuations.
The quantitative analysis on the cosmological perturbations with such new contributions is certainly important to seek observational signatures of the theory. However, this is beyond the scope of the present paper and we hope to study it in future publications.

Under the critical conditions that we have found, ghosts are absent around the flat background in the sense that their masses are infinitely heavy. In a generic curved background, the ghosts may reappear but should remain heavy as far as the deviation of the background from the flat one is not too large. In particular, it is expected that the masses of the ghosts should be roughly proportional to the inverse of the background curvature scale and that the ghosts are therefore harmless in the context of EFT as long as the curvature is small enough so that the masses of the ghosts are sufficiently heavier than the energy scale of physical processes of interest. There is also a possibility that all ghost modes can be removed even around a curved background by tuning remaining terms such as $\mathcal{L}_{RT^2}$, $\mathcal{L}_{(\nabla T)T^2}$ and $\mathcal{L}_{T^4}$. In particular, it is quite interesting if the torsionfull quadratic curvature theory can be made ghost-free at fully nonlinear order by tuning a series of higher curvature/derivative terms and re-summing them, similarly to the case of massive gravity where the de Rham-Gabadadze-Tolley (dRGT) theory provides a nonlinear completion of the Fierz-Pauli linear theory.

Theoretically, it is important to investigate the robustness of the ghost-freeness against radiative corrections. In the context of the scalar-tensor theories, it was recently found that an additional local symmetry of the connection is associated with the ghost-free structure when the spacetime geometry is extended from Riemannian one~\cite{Aoki:2018lwx,Aoki:2019rvi}. Similar considerations are certainly worthwhile in the context of higher curvature theories. Another possibility to provide the robustness is a consequence of unitarity, Lorentz invariance, causality, and locality, called the positivity bounds~\cite{Adams:2006sv}. For example, although the dRGT potential is originally discovered by just assuming the ghost-freeness up to the scale $\Lambda_3$, the positivity bounds do not accept the presence of the ghost at the scale $\Lambda_5 ~(<\Lambda_3)$ and then enforce the structure of the dRGT potential at least up to quartic order~\cite{deRham:2018qqo}. Furthermore, the recent paper \cite{Alonso:2019ptb} argues that the same assumptions predict a leading order deviation to the coupling of gravity to fermions that can be explained by the existence of the torsion. Although the present paper has used a bottom-up approach and thus has not discussed any fundamental origins of the theory, it would be interesting to investigate the connections between candidate fundamental theories of gravity and their low energy predictions about derivative corrections to the Einstein gravity in the framework beyond the Riemannian geometry. In particular, it is worthwhile to see if there are any relations between the requirements from UV theories (such as the positivity bounds) and the critical conditions summarized in Table \ref{Table1}. Yet another approach to the robustness of the ghost-freeness is to keep the masses of possible Ostrogradsky ghosts to be heavy enough. Namely, even if radiative corrections introduce failures of the critical conditions and thus reintroduce Ostrogradsky ghosts, one can still make robust predictions as far as the deviations are small enough to ensure that the masses of the Ostrogradsky ghosts are heavier than the energy scale of phenomenological interest.

In summary, we provide two new phenomenologically viable parameter spaces of the quadratic curvature theory, Class I and Class II, by means of the extension of the spacetime geometry in which not only the massive spin-0 particle and the massless graviton but also other massive spin-0,1,2 particles exist without any instability at least around the Minkowski background. It would be worth mentioning again that the present theory is not a class of PGTs since the derivatives of the torsion are explicitly included in the Lagrangian which are naturally expected from the dimensional analysis. It is thus interesting to explore phenomenological and theoretical aspects of the new type of derivative corrections to the Einstein gravity beyond the Riemannian geometry, which are left for future works.

\section*{Acknowledgments}
We would like to thank Hideo Kodama for useful comments. The work of K.A. was supported in part by Grants-in-Aid from the Scientific Research Fund of the Japan Society for the Promotion of Science  (No.~19J00895). The work of S.M. was supported by Japan Society for the Promotion of Science Grants-in-Aid for Scientific Research No. 17H02890, No. 17H06359, and by World Premier International Research Center Initiative, MEXT, Japan. 


\appendix

\section{$3+1$ decomposition}
\label{append_3+1}
Introducing a unit timelike vector $n^{\mu}$, one can define the spatial metric $\gamma_{\mu\nu}$ via
\begin{align}
\gamma_{\mu\nu}:=g_{\mu\nu}+n_{\mu}n_{\nu}.
\end{align}
The three dimensional Levi-Civita tensor $\epsilon_{\mu\nu\rho}$ is defined by
\begin{align}
\epsilon_{\mu\nu\rho}:=\epsilon_{\mu\nu\rho\sigma}n^{\sigma},
\end{align}
where $\epsilon_{\mu\nu\rho\sigma}$ is the four dimensional Levi-Civita tensor. Then, the $3+1$ components of the vector $T^{\mu}$ are
\begin{align}
\phi:=T_{\mu}n^{\mu} , \quad
A_{\alpha}:=T_{\mu}\gamma^{\mu}_{\alpha},
\end{align}
where $A_{\alpha}$ is a spatial vector, $A_{\alpha}n^{\alpha}=0$. The pseudovector $\Tcal_{\mu}$ is similarly decomposed while the $3+1$ components of $\Ta_{\mu\nu\rho}$ are defined by
\begin{align}
B_{\alpha}&:=\Ta_{\mu\nu\rho}n^{\alpha}n^{\beta}\gamma^{\rho}_{\alpha}, \quad
t_{\alpha\beta}:=\Ta_{\mu\nu\rho}\gamma^{\mu}_{(\alpha}\gamma^{\nu}_{\beta)}n^{\rho}
, \nn
\Bcal_{\alpha}&:=\frac{1}{2}\Ta_{\mu\nu\rho}n^{\mu}\epsilon^{\nu\rho}{}_{\alpha}
,\quad
\tau_{\alpha\beta}:=\frac{1}{2}\Ta_{\mu\nu\rho}\gamma^{\mu}_{(\alpha}\epsilon_{\beta)\nu\rho},
\end{align}
where $t_{\alpha\beta}$ and $\tau_{\alpha\beta}$ are symmetric and traceless by definition. Conversely, the four dimensional quantities are expressed by
\begin{align}
T_{\mu}&=-n_{\mu}\phi+A_{\mu}, \\
\Ta_{\mu\nu\rho}&=2n_{\mu}n_{[\nu}B_{\rho]}-n_{\mu}\epsilon_{\nu\rho\sigma}\Bcal^{\sigma}
-2t_{\mu[\nu}n_{\rho]}
\nn
&+\Bcal_{\sigma}\epsilon_{\mu[\nu}{}^{\sigma}n_{\rho]}+\gamma_{\mu[\nu}B_{\rho]}+\epsilon_{\nu\rho}{}^{\sigma}\tau_{\mu\sigma}.
\end{align}

\section{Ghost-free conditions and masses}
\label{append_ghost}
After expanding the spatial dependence in terms of harmonics and eliminating non-dynamical variables, the quadratic action of each SVT components is generally given by the form,
\begin{align}
S_2=\int dt \sum L, \quad L=\frac{1}{2} \left[ \dot{\qV}^T \mathcal{K} \dot{\qV}+ 2\dot{\qV}^T \mathcal{M} \qV - \qV^T \mathcal{V} \qV \right],
\end{align}
where $\qV=\{ q_i \}~(i=1,2,\cdots,N)$ is the set of dynamical variables and $\mathcal{K},\mathcal{M},\mathcal{V}$ are $N\times N$ constant matrices with ${\rm det} \mathcal{K} \neq 0, {\rm det} \mathcal{V} \neq 0$. Here, the suffix $T$ stands for transpose as usual. The sum runs over different helicity modes as well as spatial momenta. The corresponding Hamiltonian is
\begin{align}
H=\frac{1}{2}\left[ (\pV-\mathcal{M}\qV)^T \mathcal{K}^{-1} (\pV-\mathcal{M}\qV)+\qV^T \mathcal{V}\qV \right],
\end{align}
where
\begin{align}
\pV=\frac{\partial L}{\partial \dot{\qV}^T}.
\end{align}
Therefore, the stability conditions are that $\mathcal{K}$ and $\mathcal{V}$ are positive definite in high $k$ limit.
The equation of motion is
\begin{align}
\mathcal{K}\ddot{\qV}+(\mathcal{M}-\mathcal{M}^T)\dot{\qV}+\mathcal{V}\qV=0 ,
\end{align}
from which we can obtain the dispersion relation via
\begin{align}
{\rm det} [\omega^2 \mathcal{K} + i\omega (\mathcal{M}-\mathcal{M}^T)-\mathcal{V}]\propto \prod_i^N [\omega^2-(k^2+m_i^2) ]=0,
\label{det}
\end{align}
where the proportionality to the product of $\omega^2-(k^2+m_i^2)$ is a consequence of the Lorentz invariance. The tachyon free conditions are obtained from $m_i^2\geq 0$.

One may worry about the existence of the friction terms in the equations of motion, that breaks the Lorentz invariance, since the equations of motion must be given by forms of the Klein-Gordon equation. However, this is just due to the choice of the variable. By means of a change of the variables $\qV \rightarrow \mathcal{P}\qV$ where $\mathcal{P}$ is a non-degenerate matrix, the matrices can be given by following element block matrices\footnote{A systematic way is that one can first take an orthogonal transformation to diagonalize $\Kcal$ and then normalize the variables so that the elements of $\Kcal$ become $\pm 1$. Then, $\Vcal$ can be diagonalized via an indefinite orthogonal transformation. The matrix $\Mcal$ can be an upper triangular matrix via integration by part.}
\begin{align}
\Kcal=
\begin{pmatrix}
\Kcal_{11} & 0 \\
0 & \Kcal_{22}
\end{pmatrix}
,\quad
\Mcal=
\begin{pmatrix}
0 & \Mcal_{12} \\
0 & 0
\end{pmatrix}
,\quad 
\Vcal=
\begin{pmatrix}
\Vcal_{11} & 0 \\
0 & \Vcal_{22}
\end{pmatrix}
.
\end{align}
The Lagrangian is thus
\begin{align}
L&=\frac{1}{2}(\dot{\qV}_1+\Kcal_{11} \Mcal_{12} \qV_2)^T \Kcal_{11} (\dot{\qV}_1 + \Kcal_{11}^{-1} \Mcal_{12} \qV_2)
\nn&
+\frac{1}{2}\dot{\qV}_2^T \Kcal_{22} \dot{\qV}_2 -\frac{1}{2}\qV_1^T \Vcal_{11} \qV_1 -\frac{1}{2}\qV_2^T (\Vcal_{22}+\Mcal_{12}^T \Kcal_{11}^{-1} \Mcal_{12})\qV_2
\end{align}
where $\qV_1=\{ q_i \}~(i=1,2,\cdots, n)$ and $\qV_2=\{ q_i \}~(i=n+1,n+2,\cdots,N)$. We introduce new variables $\QV_1$ and write
\begin{align}
L&=\frac{1}{2}\left[ 2\QV_1^T \Kcal_{11} (\dot{\qV}_q+\Kcal_{11}\Mcal_{12}\qV_2)-\QV_1^T \Kcal_{11} \QV_1 \right]
\nn &
+\frac{1}{2}\dot{\qV}_2^T \Kcal_{22} \dot{\qV}_2 -\frac{1}{2}\qV_1^T \Vcal_{11} \qV_1 -\frac{1}{2}\qV_2^T (\Vcal_{22}+\Mcal_{12}^T \Kcal_{11}^{-1} \Mcal_{12})\qV_2 ,
\end{align}
which is equivalent to the original Lagrangian when $\QV_1$ are integrated out~\cite{DeFelice:2015moy}. We instead take integration by part to eliminate $\dot{\qV_1}$ from the Lagrangian and then integrate $\qV_1 $ out. The Lagrangian is then
\begin{align}
L&=\frac{1}{2}\dot{\QV}_1^T \Kcal_{11}\Vcal_{11}^{-1} \Kcal_{11} \dot{\QV}_1+\frac{1}{2}\dot{\qV}_2 \Kcal_{22} \dot{\qV}_2
\nn
&
-\frac{1}{2}(\QV_1 - \Kcal_{11}^{-1}\Mcal_{12}\qV_2)^T \Kcal_{11}(\QV_1-\Kcal_{11}^{-1}\Mcal_{12}\qV_2)
\nn
&-\frac{1}{2}\qV_2^T \Vcal_{22} \qV_2,
\end{align}
which does not yield the friction term. The variables can be further changed $\QV:=\{\QV_1,\qV_2 \} \rightarrow \mathcal{P}'\QV$ so that both the kinetic matrix and the mass matrix are diagonalized. Since we are studying a Lorentz invariant theory, the final Lagrangian must be 
\begin{align}
L=\frac{1}{2}\dot{\QV}^T \mathcal{I} \dot{\QV}-\frac{1}{2}k^2\QV^T \mathcal{I} \QV-\frac{1}{2}\QV^T M^T \mathcal{I} M\QV
\label{L_after}
\end{align}
where $\mathcal{I}={\rm diag}(\lambda_1,\lambda_2,\cdots,\lambda_N)$ with $\lambda_i=\pm 1$ and $M={\rm diag}(m_1,m_2,\cdots,m_N)$. The ghost-freeness of the theory is the positiveness of $\lambda_i$ and the masses of the particles are $m_i$. However, this requires cumbersome calculations. The easiest way to evaluate the ghost-freeness and the masses is, as explained before, just to check the positiveness of $\Kcal$ and $\Vcal$ in high $k$ limit and to compute the nodes of the determinant \eqref{det}.

Note that only the positiveness of $\Kcal$ does not guarantee the ghost-freeness. This is because one needs to change the variables to obtain the Klein-Gordon form \eqref{L_after} after which the sign of $\lambda_i$ depends on $\Vcal$ as well. The positiveness of the kinetic matrix of \eqref{L_after} is guaranteed by the positiveness of not only $\Kcal$ but also $\Vcal$.

\section{A toy model of degenerate theory}
\label{append_toy}
We study a toy model
\begin{align}
L=\frac{1}{2} \dot{\phi}^2+\frac{c_1}{2} \ddot{\phi}^2 +c_2 \ddot{\phi}\dot{q} +\frac{c_3}{2} \dot{q}^2 -\frac{1}{2}m^2 q^2,
\end{align}
in classical mechanics. We introduce a new variable $\Phi$ and reduce the Lagrangian into the first order system
\begin{align}
L=\frac{1}{2} \dot{\phi}^2+\frac{c_1}{2} \Phi^2 + c_2 \Phi \dot{q} + \frac{c_3}{2} \dot{q}^2 + \lambda(\ddot{\phi}-\Phi)
-\frac{1}{2}m^2 q^2. \label{L1}
\end{align}
Since $\Phi$ is non-dynamical, it can be eliminated by the use of its equation of motion. We then obtain
\begin{align}
L&=\frac{1}{2} \dot{\phi}^2 -\dot{\phi}\dot{\lambda}  +\frac{1}{2c_1}(c_1 c_3- c_2^2) \dot{q}^2 
\nn
&+ \frac{c_2}{c_1}\lambda \dot{q} - \frac{1}{2c_1}\lambda^2 -\frac{1}{2}m^2 q^2 . \label{Lghost}
\end{align}
The kinetic matrix always has a negative eigenvalue and thus there exists a ghost. However, if the Lagrangian is degenerate, that is if $c_2=\sqrt{c_1c_3}$, then the variable $q$ becomes non-dynamical and then solving it gives
\begin{align}
L=\frac{1}{2}\dot{\phi}^2-\dot{\phi}\dot{\lambda}+\frac{c_3}{2c_1 m^2} \dot{\lambda}^2 - \frac{\lambda^2}{2c_1}, \label{L2}
\end{align}
which is free from ghost and tachyon when
\begin{align}
c_3>0 ,~ \frac{c_3}{m^2}>c_1>0 . \label{cond}
\end{align}
This is the idea of the degenerate higher order theories~\cite{Langlois:2015cwa}. Note that the kinetic matrix of \eqref{Lghost} has at least one negative eigenvalue for any choice of parameters $c_i$. Nonetheless, the integrating-out of $q$ changes the kinetic matrix of the Lagrangian due to the coupling $\lambda \dot{q}$ and then all eigenvalues of the kinetic matrix of \eqref{L2} can be positive.

The degeneracy condition indeed corresponds to the condition that the mass of ghost becomes infinity. When the parameter is assumed to be
\begin{align}
c_2=\sqrt{c_1c_3}(1+\epsilon ),
\end{align}
with $\epsilon \ll 1$, the masses of the normal mode and the ghost mode are
\begin{align}
m_{\rm normal}^2&=\frac{m^2}{2(c_3-c_1m^2)} +O(\epsilon^1), \nn
m^2_{\rm ghost}&= \frac{c_3-c_1m^2}{4c_1c_3} \frac{1}{\epsilon} +O(\epsilon^0) ,
\end{align}
respectively. Therefore, even if the degeneracy condition holds only approximately but not exactly, one can obtain a stable theory below a cut-off. 

\bibliography{ref}

\providecommand{\href}[2]{#2}\begingroup\raggedright\begin{thebibliography}{10}

\bibitem{Starobinsky:1980te}
A.~A. Starobinsky, \emph{{A New Type of Isotropic Cosmological Models Without
  Singularity}},
  \href{http://dx.doi.org/10.1016/0370-2693(80)90670-X}{\emph{Phys. Lett.} {\bf
  B91} (1980) 99--102}.

\bibitem{Akrami:2018odb}
{\scshape Planck} collaboration, Y.~Akrami et~al., \emph{{Planck 2018 results.
  X. Constraints on inflation}},  \href{http://arxiv.org/abs/1807.06211}{{\tt
  1807.06211}}.

\bibitem{Hazumi:2019lys}
M.~Hazumi et~al., \emph{{LiteBIRD: A Satellite for the Studies of B-Mode
  Polarization and Inflation from Cosmic Background Radiation Detection}},
  \href{http://dx.doi.org/10.1007/s10909-019-02150-5}{\emph{J. Low. Temp.
  Phys.} {\bf 194} (2019) 443--452}.

\bibitem{Fujii:2003pa}
Y.~Fujii and K.~Maeda, \emph{{The scalar-tensor theory of gravitation}}.
\newblock Cambridge University Press, 2007.

\bibitem{Boulware:1973my}
D.~G. Boulware and S.~Deser, \emph{{Can gravitation have a finite range?}},
  \href{http://dx.doi.org/10.1103/PhysRevD.6.3368}{\emph{Phys. Rev.} {\bf D6}
  (1972) 3368--3382}.

\bibitem{deRham:2010ik}
C.~de~Rham and G.~Gabadadze, \emph{{Generalization of the Fierz-Pauli Action}},
  \href{http://dx.doi.org/10.1103/PhysRevD.82.044020}{\emph{Phys. Rev.} {\bf
  D82} (2010) 044020}, [\href{http://arxiv.org/abs/1007.0443}{{\tt
  1007.0443}}].

\bibitem{deRham:2010kj}
C.~de~Rham, G.~Gabadadze and A.~J. Tolley, \emph{{Resummation of Massive
  Gravity}},
  \href{http://dx.doi.org/10.1103/PhysRevLett.106.231101}{\emph{Phys. Rev.
  Lett.} {\bf 106} (2011) 231101}, [\href{http://arxiv.org/abs/1011.1232}{{\tt
  1011.1232}}].

\bibitem{Gleyzes:2014dya}
J.~Gleyzes, D.~Langlois, F.~Piazza and F.~Vernizzi, \emph{{Healthy theories
  beyond Horndeski}},
  \href{http://dx.doi.org/10.1103/PhysRevLett.114.211101}{\emph{Phys. Rev.
  Lett.} {\bf 114} (2015) 211101}, [\href{http://arxiv.org/abs/1404.6495}{{\tt
  1404.6495}}].

\bibitem{Gleyzes:2014qga}
J.~Gleyzes, D.~Langlois, F.~Piazza and F.~Vernizzi, \emph{{Exploring
  gravitational theories beyond Horndeski}},
  \href{http://dx.doi.org/10.1088/1475-7516/2015/02/018}{\emph{JCAP} {\bf 1502}
  (2015) 018}, [\href{http://arxiv.org/abs/1408.1952}{{\tt 1408.1952}}].

\bibitem{Zumalacarregui:2013pma}
M.~Zumalac{\'a}rregui and J.~Garc{\'\i}a-Bellido, \emph{{Transforming gravity:
  from derivative couplings to matter to second-order scalar-tensor theories
  beyond the Horndeski Lagrangian}},
  \href{http://dx.doi.org/10.1103/PhysRevD.89.064046}{\emph{Phys. Rev.} {\bf
  D89} (2014) 064046}, [\href{http://arxiv.org/abs/1308.4685}{{\tt
  1308.4685}}].

\bibitem{Langlois:2015cwa}
D.~Langlois and K.~Noui, \emph{{Degenerate higher derivative theories beyond
  Horndeski: evading the Ostrogradski instability}},
  \href{http://dx.doi.org/10.1088/1475-7516/2016/02/034}{\emph{JCAP} {\bf 1602}
  (2016) 034}, [\href{http://arxiv.org/abs/1510.06930}{{\tt 1510.06930}}].

\bibitem{Langlois:2015skt}
D.~Langlois and K.~Noui, \emph{{Hamiltonian analysis of higher derivative
  scalar-tensor theories}},
  \href{http://dx.doi.org/10.1088/1475-7516/2016/07/016}{\emph{JCAP} {\bf 1607}
  (2016) 016}, [\href{http://arxiv.org/abs/1512.06820}{{\tt 1512.06820}}].

\bibitem{Crisostomi:2016czh}
M.~Crisostomi, K.~Koyama and G.~Tasinato, \emph{{Extended Scalar-Tensor
  Theories of Gravity}},
  \href{http://dx.doi.org/10.1088/1475-7516/2016/04/044}{\emph{JCAP} {\bf 1604}
  (2016) 044}, [\href{http://arxiv.org/abs/1602.03119}{{\tt 1602.03119}}].

\bibitem{Achour:2016rkg}
J.~Ben~Achour, D.~Langlois and K.~Noui, \emph{{Degenerate higher order
  scalar-tensor theories beyond Horndeski and disformal transformations}},
  \href{http://dx.doi.org/10.1103/PhysRevD.93.124005}{\emph{Phys. Rev.} {\bf
  D93} (2016) 124005}, [\href{http://arxiv.org/abs/1602.08398}{{\tt
  1602.08398}}].

\bibitem{BenAchour:2016fzp}
J.~Ben~Achour, M.~Crisostomi, K.~Koyama, D.~Langlois, K.~Noui and G.~Tasinato,
  \emph{{Degenerate higher order scalar-tensor theories beyond Horndeski up to
  cubic order}}, \href{http://dx.doi.org/10.1007/JHEP12(2016)100}{\emph{JHEP}
  {\bf 12} (2016) 100}, [\href{http://arxiv.org/abs/1608.08135}{{\tt
  1608.08135}}].

\bibitem{Will:2014kxa}
C.~M. Will, \emph{{The Confrontation between General Relativity and
  Experiment}}, \href{http://dx.doi.org/10.12942/lrr-2014-4}{\emph{Living Rev.
  Rel.} {\bf 17} (2014) 4}, [\href{http://arxiv.org/abs/1403.7377}{{\tt
  1403.7377}}].

\bibitem{Percacci:2009ij}
R.~Percacci, \emph{{Gravity from a Particle Physicists' perspective}},
  \href{http://dx.doi.org/10.22323/1.081.0011}{\emph{PoS} {\bf ISFTG} (2009)
  011}, [\href{http://arxiv.org/abs/0910.5167}{{\tt 0910.5167}}].

\bibitem{Blagojevic:2002du}
M.~Blagojevic, \emph{{Gravitation and gauge symmetries}}.
\newblock 2002.

\bibitem{Obukhov:2018bmf}
Y.~N. Obukhov, \emph{{Poincar\'{e} gauge gravity: An overview}},
  \href{http://dx.doi.org/10.1142/S0219887818400054}{\emph{Int. J. Geom. Meth.
  Mod. Phys.} {\bf 15} (2018) 1840005},
  [\href{http://arxiv.org/abs/1805.07385}{{\tt 1805.07385}}].

\bibitem{Sezgin:1979zf}
E.~Sezgin and P.~van Nieuwenhuizen, \emph{{New Ghost Free Gravity Lagrangians
  with Propagating Torsion}},
  \href{http://dx.doi.org/10.1103/PhysRevD.21.3269}{\emph{Phys. Rev.} {\bf D21}
  (1980) 3269}.

\bibitem{Lin:2018awc}
Y.-C. Lin, M.~P. Hobson and A.~N. Lasenby, \emph{{Ghost and tachyon free
  Poincar\'{e} gauge theories: A systematic approach}},
  \href{http://dx.doi.org/10.1103/PhysRevD.99.064001}{\emph{Phys. Rev.} {\bf
  D99} (2019) 064001}, [\href{http://arxiv.org/abs/1812.02675}{{\tt
  1812.02675}}].

\bibitem{Alvarez-Gaume:2015rwa}
L.~Alvarez-Gaume, A.~Kehagias, C.~Kounnas, D.~L{\"u}st and A.~Riotto,
  \emph{{Aspects of Quadratic Gravity}},
  \href{http://dx.doi.org/10.1002/prop.201500100}{\emph{Fortsch. Phys.} {\bf
  64} (2016) 176--189}, [\href{http://arxiv.org/abs/1505.07657}{{\tt
  1505.07657}}].

\bibitem{Stelle:1976gc}
K.~S. Stelle, \emph{{Renormalization of Higher Derivative Quantum Gravity}},
  \href{http://dx.doi.org/10.1103/PhysRevD.16.953}{\emph{Phys. Rev.} {\bf D16}
  (1977) 953--969}.

\bibitem{Mukohyama:2013ew}
S.~Mukohyama and J.-P. Uzan, \emph{{From configuration to dynamics: Emergence
  of Lorentz signature in classical field theory}},
  \href{http://dx.doi.org/10.1103/PhysRevD.87.065020}{\emph{Phys. Rev.} {\bf
  D87} (2013) 065020}, [\href{http://arxiv.org/abs/1301.1361}{{\tt
  1301.1361}}].

\bibitem{Mukohyama:2013gra}
S.~Mukohyama, \emph{{Emergence of time in power-counting renormalizable
  Riemannian theory of gravity}},
  \href{http://dx.doi.org/10.1103/PhysRevD.87.085030}{\emph{Phys. Rev.} {\bf
  D87} (2013) 085030}, [\href{http://arxiv.org/abs/1303.1409}{{\tt
  1303.1409}}].

\bibitem{Anselmi:2017ygm}
D.~Anselmi, \emph{{On the quantum field theory of the gravitational
  interactions}}, \href{http://dx.doi.org/10.1007/JHEP06(2017)086}{\emph{JHEP}
  {\bf 06} (2017) 086}, [\href{http://arxiv.org/abs/1704.07728}{{\tt
  1704.07728}}].

\bibitem{deRham:2014zqa}
C.~de~Rham, \emph{{Massive Gravity}},
  \href{http://dx.doi.org/10.12942/lrr-2014-7}{\emph{Living Rev. Rel.} {\bf 17}
  (2014) 7}, [\href{http://arxiv.org/abs/1401.4173}{{\tt 1401.4173}}].

\bibitem{Schmidt-May:2015vnx}
A.~Schmidt-May and M.~von Strauss, \emph{{Recent developments in bimetric
  theory}}, \href{http://dx.doi.org/10.1088/1751-8113/49/18/183001}{\emph{J.
  Phys.} {\bf A49} (2016) 183001}, [\href{http://arxiv.org/abs/1512.00021}{{\tt
  1512.00021}}].

\bibitem{Hayashi:1979wj}
K.~Hayashi and T.~Shirafuji, \emph{{Gravity from Poincare Gauge Theory of the
  Fundamental Particles. 1. Linear and Quadratic Lagrangians}},
  \href{http://dx.doi.org/10.1143/PTP.64.866}{\emph{Prog. Theor. Phys.} {\bf
  64} (1980) 866}.

\bibitem{Hayashi:1980ir}
K.~Hayashi and T.~Shirafuji, \emph{{Gravity From Poincare Gauge Theory of the
  Fundamental Particles. 3. Weak Field Approximation}},
  \href{http://dx.doi.org/10.1143/PTP.64.1435}{\emph{Prog. Theor. Phys.} {\bf
  64} (1980) 1435}.

\bibitem{Hayashi:1980qp}
K.~Hayashi and T.~Shirafuji, \emph{{Gravity From Poincare Gauge Theory of the
  Fundamental Particles. 4. Mass and Energy of Particle Spectrum}},
  \href{http://dx.doi.org/10.1143/PTP.64.2222}{\emph{Prog. Theor. Phys.} {\bf
  64} (1980) 2222}.

\bibitem{Nair:2008yh}
V.~P. Nair, S.~Randjbar-Daemi and V.~Rubakov, \emph{{Massive Spin-2 fields of
  Geometric Origin in Curved Spacetimes}},
  \href{http://dx.doi.org/10.1103/PhysRevD.80.104031}{\emph{Phys. Rev.} {\bf
  D80} (2009) 104031}, [\href{http://arxiv.org/abs/0811.3781}{{\tt
  0811.3781}}].

\bibitem{Nikiforova:2009qr}
V.~Nikiforova, S.~Randjbar-Daemi and V.~Rubakov, \emph{{Infrared Modified
  Gravity with Dynamical Torsion}},
  \href{http://dx.doi.org/10.1103/PhysRevD.80.124050}{\emph{Phys. Rev.} {\bf
  D80} (2009) 124050}, [\href{http://arxiv.org/abs/0905.3732}{{\tt
  0905.3732}}].

\bibitem{Arkani-Hamed:2015bza}
N.~Arkani-Hamed and J.~Maldacena, \emph{{Cosmological Collider Physics}},
  \href{http://arxiv.org/abs/1503.08043}{{\tt 1503.08043}}.

\bibitem{Aoki:2018lwx}
K.~Aoki and K.~Shimada, \emph{{Galileon and generalized Galileon with
  projective invariance in a metric-affine formalism}},
  \href{http://dx.doi.org/10.1103/PhysRevD.98.044038}{\emph{Phys. Rev.} {\bf
  D98} (2018) 044038}, [\href{http://arxiv.org/abs/1806.02589}{{\tt
  1806.02589}}].

\bibitem{Aoki:2019rvi}
K.~Aoki and K.~Shimada, \emph{{Scalar-metric-affine theories: Can we get
  ghost-free theories from symmetry?}},
  \href{http://dx.doi.org/10.1103/PhysRevD.100.044037}{\emph{Phys. Rev.} {\bf
  D100} (2019) 044037}, [\href{http://arxiv.org/abs/1904.10175}{{\tt
  1904.10175}}].

\bibitem{Adams:2006sv}
A.~Adams, N.~Arkani-Hamed, S.~Dubovsky, A.~Nicolis and R.~Rattazzi,
  \emph{{Causality, analyticity and an IR obstruction to UV completion}},
  \href{http://dx.doi.org/10.1088/1126-6708/2006/10/014}{\emph{JHEP} {\bf 10}
  (2006) 014}, [\href{http://arxiv.org/abs/hep-th/0602178}{{\tt
  hep-th/0602178}}].

\bibitem{deRham:2018qqo}
C.~de~Rham, S.~Melville, A.~J. Tolley and S.-Y. Zhou, \emph{{Positivity Bounds
  for Massive Spin-1 and Spin-2 Fields}},
  \href{http://dx.doi.org/10.1007/JHEP03(2019)182}{\emph{JHEP} {\bf 03} (2019)
  182}, [\href{http://arxiv.org/abs/1804.10624}{{\tt 1804.10624}}].

\bibitem{Alonso:2019ptb}
R.~Alonso and A.~Urbano, \emph{{On amplitudes, resonances and the ultraviolet
  completion of gravity}},  \href{http://arxiv.org/abs/1906.11687}{{\tt
  1906.11687}}.

\bibitem{DeFelice:2015moy}
A.~De~Felice and S.~Mukohyama, \emph{{Phenomenology in minimal theory of
  massive gravity}},
  \href{http://dx.doi.org/10.1088/1475-7516/2016/04/028}{\emph{JCAP} {\bf 1604}
  (2016) 028}, [\href{http://arxiv.org/abs/1512.04008}{{\tt 1512.04008}}].

\end{thebibliography}\endgroup
\bibliographystyle{JHEP}

\end{document}